\documentclass[prb,twocolumn,aps,showpacs,superscriptaddress,epsfig]{revtex4-2}

\usepackage[utf8]{inputenc}
\usepackage{amsmath,amssymb,amsfonts}
\usepackage{bm}
\usepackage{physics} 

\usepackage{graphicx}
\usepackage{subcaption} 
\usepackage{float}
\usepackage[export]{adjustbox} 

\usepackage{dcolumn}

\usepackage{lipsum} 
\usepackage{soul} 
\usepackage{mdframed} 
\usepackage{ragged2e} 

\usepackage{mathrsfs} 

\usepackage{appendix}
\usepackage{caption} 

\usepackage{tikz} 

\usepackage{hyperref}
\hypersetup{
    colorlinks=true,
    linkcolor=blue,
    citecolor = blue,
    filecolor=magenta,      
    urlcolor=black,
    pdftitle={Random_circ_TR}
}

\newcommand\Tau{\mathcal{T}}

\definecolor{OliveGreen}{cmyk}{0.64, 0, 0.95, 0.40}

\newcommand{\an}[1]{\textcolor{OliveGreen}{#1}}

\newcommand*{\rom}[1]{\expandafter\@slowromancap\romannumeral #1@}

\begin{document}
\preprint{APS/123-QED}
\title{Random Quantum Circuits with Time-Reversal Symmetry}

\author{Kabir Khanna}
\affiliation{Department of Physics, University of Massachusetts, Amherst, Massachusetts 01003, USA}
\author{Abhishek Kumar}
\affiliation{Department of Physics, University of Massachusetts, Amherst, Massachusetts 01003, USA}

\author{Romain Vasseur}%
 \email{romain.vasseur@unige.ch}
 \affiliation{Department of Theoretical Physics, University of Geneva, 24 quai Ernest-Ansermet, 1211 Gen\`eve, Switzerland}
\affiliation{Department of Physics, University of Massachusetts, Amherst, Massachusetts 01003, USA}

\author{Andreas~W.~W.~Ludwig}
\affiliation{Department of Physics, University of California, Santa Barbara, California 93106, USA}

\begin{abstract}
Time-reversal (TR) symmetry is crucial for understanding a wide range of physical phenomena, and plays a key role in constraining fundamental particle interactions and in classifying phases of quantum matter. In this work, we introduce an ensemble of random quantum circuits that are representative of the dynamics of generic TR-invariant many-body quantum systems. We derive a general statistical mechanics model describing entanglement, many-body quantum chaos and quantum information dynamics in such TR-invariant circuits.
As an example of application of our formalism, we study the universal properties of measurement-induced phase transitions (MIPT) in monitored TR-invariant systems, with measurements performed in a TR-invariant basis. We find that TR-invariance of the unitary part of the dynamics does not affect the universality class, unless measurement outcomes are post-selected to satisfy the global TR-invariance of each quantum trajectory. We confirm these predictions numerically, and find, for both generic and Clifford-based evolutions, novel critical exponents in the case of ``strong'', i.e.~global TR-invariance where each quantum trajectory is TR-invariant. \\

\end{abstract}

\maketitle
\tableofcontents


\section{Introduction}
Generic isolated many-body systems undergoing unitary dynamics thermalize by scrambling locally encoded information into non-local degrees of freedom~\cite{Deutsch1991,Srednicki1994, Rigol2008, RevModPhys.91.021001}. In contrast, open quantum systems can be modeled by two competing processes: unitary evolution, which scrambles information, and non-unitary generalized measurements due to noisy coupling with the environment, which attempt to extract quantum information from the system. Understanding the universal features of this competition presents a key challenge. Recent experimental progress with digital and noisy-intermediate scale quantum simulators has further accelerated theoretical advancements in this area~\cite{Georgescu2014, Preskill2018}. Minimally structured models that can naturally be implemented on these platforms and capture the relevant physics include quantum circuits with random local unitary gates and measurements. The randomness in these circuits often allows for theoretical tractability by mapping specific problems to effective classical descriptions. Consequently, these systems have been instrumental in uncovering many universal phenomena associated with the dynamics of quantum information, even in the absence of symmetries. This includes KPZ universality~\cite{KardarParisiZhang1986} of entanglement growth~\cite{PhysRevX.7.031016, PhysRevB.99.174205}, its connection to the Ryu-Takayanagi formula in holography~\cite{RyuTakayanagi2006, RyuTakayanagi2006b}, and the hydrodynamical nature of operator spreading~\cite{PhysRevX.8.021013, PhysRevX.8.021014} among others.

By adding local quantum measurements to the dynamics, an intriguing dynamical phase transition appears in the quantum trajectories as the measurement rate increases, known as a measurement-induced phase transition (MIPT) ~\cite{PhysRevX.9.031009, PhysRevB.98.205136}. Crucially, this transition is only visible in properties of a quantum state {\em conditional } on a set of measurement outcomes -- it is invisible if measurement outcomes are discarded (or traced over). 
The MIPT phases are distinguished by the entanglement properties of quantum trajectories: at low measurement rates, entanglement follows a volume law, while at high rates, it obeys an area law. Interestingly, this transition can be understood from multiple perspectives. One such perspective is the purification transition, where a mixed state remains mixed in the volume law phase but is purified in the area law phase~\cite{PhysRevX.10.041020}. Another viewpoint comes from the lens of quantum error correction: the volume law phase serves as an efficient error-correcting code, encoding information non-locally and safeguarding it from local errors (measurements)~\cite{PhysRevLett.125.030505}. Lastly, this transition can also be understood from the perspective of the observer, in terms of ``learnability" of the initial state~\cite{PhysRevLett.129.200602, PhysRevLett.125.030505, PRXQuantum.5.020304, PhysRevX.10.041020}. In the area law phase, measurement outcomes contain enough information to help distinguish between two states, whereas in the volume law phase, the outcomes 
are essentially random, making it impossible to infer the difference between them. 

MIPTs have been studied extensively in a variety of contexts both numerically and theoretically in the past few years, establishing them as a generic property of monitored quantum systems \cite{PhysRevX.10.041020, PhysRevLett.125.030505, PhysRevB.103.104306, PhysRevB.103.174309, PhysRevB.100.134306, PhysRevB.99.224307, PhysRevB.104.104305, 10.21468/SciPostPhys.7.2.024, PhysRevB.100.064204, PhysRevLett.125.070606, PhysRevB.101.060301, PhysRevResearch.2.023288, PhysRevX.11.011030, lavasani_measurement-induced_2021, PhysRevResearch.3.023200, PhysRevResearch.2.013022, PhysRevB.102.014315, PhysRevB.102.054302, PhysRevLett.129.200602, PhysRevB.104.155111, PhysRevB.103.224210, PhysRevLett.126.060501,PRXQuantum.2.040319, PhysRevB.106.134206, PhysRevLett.126.170503, PhysRevB.106.144313, BAO2021168618, PhysRevLett.128.010604, PhysRevB.104.094304, PhysRevLett.128.050602, PhysRevX.12.041002, PhysRevB.108.214302,PhysRevLett.126.170602, jian2021quantumerroremergentmagnetic, PhysRevResearch.4.023146, PhysRevLett.128.130605,PhysRevX.11.041004, Sierant2022dissipativefloquet, 10.21468/SciPostPhysCore.5.2.023, PhysRevLett.130.220404, PhysRevB.109.125148, PhysRevResearch.6.033220, PRXQuantum.4.030333, PhysRevB.110.064309, PhysRevB.110.064301, PhysRevB.110.L060202, PhysRevB.108.184302, PhysRevB.107.L220204, PRXQuantum.5.020304, PhysRevB.108.184204, PhysRevB.109.014303, PhysRevB.107.L201113, PRXQuantum.5.030329, PhysRevB.110.054308, PhysRevResearch.6.023176, PhysRevLett.130.120402, PhysRevLett.132.240402, PhysRevLett.130.230401, PhysRevB.107.L220201, PhysRevB.110.045135, PhysRevB.108.L041103, PhysRevB.107.094309,PhysRevB.110.064323,PhysRevB.107.014308,PhysRevB.107.224303,PhysRevLett.131.060403,PhysRevLett.131.220404,PhysRevB.108.104310,PhysRevB.108.165126,PhysRevB.108.104203,PhysRevB.107.064303, Li_2023,PhysRevB.108.L020306,JianShapourianBauerLudwig2023,
PhysRevX.13.041045,PhysRevX.13.041046,PRXQuantum.4.040332} (also see~\cite{Potter_2022,annurev:/content/journals/10.1146/annurev-conmatphys-031720-030658} for reviews on the subject). The credibility of these studies is further supported by experimental observations of MIPTs in numerous experimental setups \cite{noel_measurement-induced_2022, koh_measurement-induced_2023, hoke_measurement-induced_2023, agrawal2023observingquantummeasurementcollapse, kamakari2024experimentaldemonstrationscalablecrossentropy}. A particularly important theoretical tool developed to understand and characterize these transitions has been to map them into effective stat-mech models using the replica trick, a procedure that was first outlined in the study of random tensor networks \cite{hayden_holographic_2016, PhysRevB.100.134203} and was subsequently adopted for unitary and hybrid circuits 
\cite{PhysRevB.99.174205, PhysRevB.101.104301, PhysRevB.101.104302, PhysRevB.109.174307}.
The need for the replica trick here emerges when dealing with observables that are inherently non-linear in the density matrix. An example of the utility of these stat-mech mappings is the understanding of the entanglement transition as a symmetry breaking transition. Here the volume law corresponds to the symmetry-broken phase, while the area law reflects a disordered phase. Entanglement scaling, say in the volume law phase, is then understood as resulting from the domain wall line tension at the boundary.

An important area of research within random quantum circuits has focused on using these tools to explore the effects of introducing global symmetry into the system. Previous studies on Abelian symmetries such as U(1) and non-Abelian symmetries such as SU(2) have revealed interesting physics~\cite{PhysRevLett.123.210603,  PhysRevX.8.031058, PhysRevX.8.031057, PhysRevB.108.054307, PhysRevLett.131.210402}. Notably, there can exist a new ``charge sharpening'' transition, where each phase differs based on how effectively an observer can learn about global quantum numbers from local measurements~\cite{PhysRevX.12.041002,PhysRevB.108.054307,GuoJianFosterLudwigKeldysh2024,MirlinGornyiEtAlKeldysh2024}.

In contrast to these internal symmetries, the role of fundamental symmetries such as
time-reversal (TR)
on the dynamics of quantum information remains relatively unexplored. 
Historically, this approach underpinned Dyson's three-fold classification of random matrix ensembles, of which TR (``orthogonal" ensemble) was a central class  ~\cite{Dyson1962Statistical}. More recently, Zirnbauer \cite{zirnbauer1996riemannian} and Altland and Zirnbauer \cite{PhysRevB.55.1142} enumerated all possible symmetry classes of random matrices 
and of Hamiltonians of non-interacting fermionic systems.
There were ten classes in total, including the three classes Dyson introduced. 
 TR symmetry was again used here as one of a complete set of anti-unitary 
symmetries which enable an exhaustive classification of all Hamiltonians modulo unitarily implemented symmetries~\cite{Heinzner_2005}~\footnote{For a basic review see, e.g., \cite{LudwigNobelSymposium}}.
This 
exhaustive list of symmetry classes
then culminated in
the  ``ten-fold way" 
classification 
of topological insulators and superconductors for 
non-interacting fermions~\cite{ PhysRevB.78.195125,Ryu2010Topological}. Finally, TR-symmetry also played a crucial role in the classification of topological phases in interacting systems
(see, e.g., \cite{PhysRevB.83.075103,ChongSenthilPRB2014}).
For an approach, which is different from the logic presented in the present paper, using the "10-fold way" for the classification of non-unitary circuit evolutions of non-interacting fermions, see \cite{PhysRevB.106.134206}.
For related work in a somewhat different context, see \cite{PanShapourianJianTopological2024}.
Given such historical significance of TR, in this paper we explore the dynamics of TR-invariant chaotic quantum systems 
where the (anti-unitary) time reversal operator $\Tau$ squares to 
plus the identity,
$\Tau^2= + \mathrm{1}$ (``orthogonal class''),
using the toolbox of random quantum circuits.
(In this paper we only consider for simplicity this
particular  time-reversal  symmetry, referring to it simply as TR symmetry from now on. Circuits invariant under the time-reversal operation where $\Tau^2= - \mathrm{1}$ will be treated in future work.)
As we review in section~\ref{trsection}, imposing TR-invariance leads to selecting gates from the Circular Orthogonal Ensemble (COE), 
which consists of all symmetric unitary matrices and is thus
invariant under $U\rightarrow \mathcal{U}^TU\mathcal{U}$, where $\mathcal{U}$ is a fixed unitary. 
We make a crucial distinction between ``local TR invariance" and ``global TR invariance", notions we detail in section~\ref{trsection}. Essentially, local TR invariance refers to models where each gate used in the evolution is sampled from the COE, and is naturally generated by a TR-invariant Hamiltonian. There is a natural reason to study such local TR setups, namely, that any evolution generated by the Trotterized version of a TR invariant Hamiltonian would have local TR invariance. Global TR invariance on the other hand puts restriction on the whole evolution 
of the entire circuit being TR symmetric.

The key result of our work is a general mapping of the calculation of the dynamics of Rényi entropies in TR-invariant quantum circuits onto a replica statistical mechanics model. This mapping generalizes the approach of Refs.~\cite{PhysRevB.99.174205, PhysRevB.101.104301, PhysRevB.101.104302}, but involves averaging over moments of COE matrices
 instead of over unitary matrices, performed graphically using results from
 the corresponding random matrix theory detailed in~\cite{matsumotogeneral}. The resulting statistical mechanics model features non-local Boltzmann weights, differing qualitatively from the familiar statistical mechanics model of the canonical
 unitary Haar ensemble.  This mapping provides the first step towards studying properties of many-body chaos under TR-invariant unitary 
 evolution using the tools of random quantum quantum circuits,
 including, e.g.,  operator spreading and entanglement dynamics.

As an example of application of this framework, we then study MIPTs in monitored TR-invariant quantum dynamics. Here, the measurements are performed in a TR-invariant basis, where the measurement outcomes ``on average" preserve the TR-invariance of the dynamics. One might then expect a novel universality class to emerge. However, we find that the universality class of the MIPT remains the same as in the generic, non TR-symmetric model, as random measurement outcomes break TR-invariance in individual trajectories. On the other hand, if we post-select measurement outcomes so each quantum trajectory is individually TR-invariant (``strong'' global symmetry), we predict that a new universality class emerges. We interpret these results in terms of the symmetry of the underlying statistical mechanics models.
We note that in a ``folded'' formulation of the circuit with ``strong'' global TR symmetry, the post-selection requirement of measurement outcomes is replaced by working with the 2nd tensor power of the evolution operator of the first half of the circuit in a doubled Hilbert space and no post-selection (Sec.~\ref{globaltrmipt}).
Our findings are supported by numerical results from both Haar and Clifford versions of these circuits, identifying the critical points and distinguishing the aforementioned universality classes. Finally, we note that there have been previous works that have studied random quantum circuits with TR by sampling the gates from COE \cite{hunterjones2018operatorgrowthrandomquantum, Kalsi_2022}. However, those works do not discuss any critical features of the measurement-induced transition. More importantly, they fail to address the key distinction between global and local time-reversal symmetry, a topic that forms a significant part of our discussion. Finally, we note that the notion of TR discussed in our work arises essentially from requiring a solution to the Schrödinger equation invariant under the transformation \( t \rightarrow -t \). This should not be confused with the fundamentally different setup of a literal reversal of dynamics generated by
 a unitary \( U \) through the subsequent application of
 the unitary \( U^\dagger \), 
as discussed, e.g., in
\cite{khindanov2024observable, PhysRevLett.132.140401, niroula_phase_2024}.

The plan for the remainder of this work is as follows: In Section \ref{trsection}, we discuss TR symmetry, distinguishing between local and global TR. Section \ref{smmodels} presents the stat-mech mappings for both the local and global TR model, 
for purely unitary evolution, i.e. in the absence of measurements. Section \ref{miptsection} applies these models to analyze MIPTs, discussing symmetry properties, universality, and the analytically tractable 
limit of large on-site Hilbert space dimension $d$. Finally, Section \ref{numerics} presents numerical results for both Haar and Clifford circuits. We end with section \ref{outlook} where we summarize our findings and discuss their broader implications.


\section{Time-reversal invariant quantum dynamics}
\label{trsection}

\begin{figure}
    \centering
    \includegraphics[width=0.8\linewidth, trim={0.52cm 0 0.52cm 0}, clip]{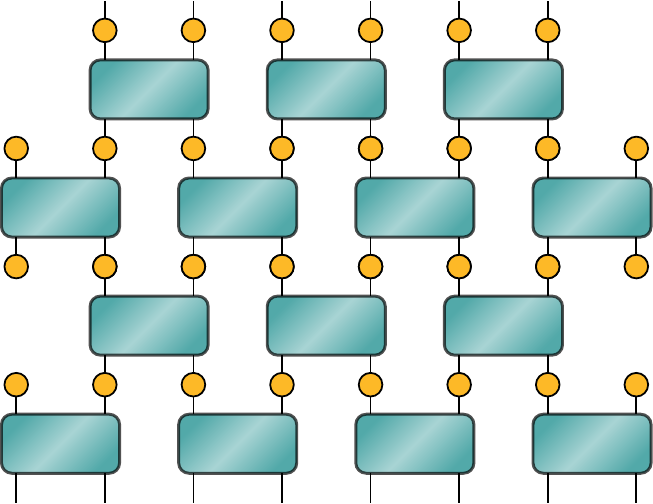}
        \captionsetup{justification=Justified, singlelinecheck=false, font=small}
    \caption{{\bf Monitored TR-invariant system.} 
    Physical setup considered in this paper of a weakly monitored TR-invariant one-dimensional quantum system. Upon Trotterization, we obtain a quantum circuit with symmetric gates $U^T=U$, which are all identical as indicated by their color. }
    \label{fig:enter-label}
\end{figure}

We consider a one-dimensional many-body quantum system of qudits with Hilbert space ${\cal H} = (\mathbb{C}^{d})^{\otimes L}$. We are broadly interested in the dynamics of quantum information in the case where this system is invariant under time-reversal. More generally, we will also consider monitored TR-invariant systems -- where the randomness of the measurement outcomes breaks TR invariance in each quantum trajectory, but where the unitary part of the dynamics is TR-invariant. 

\begin{figure*}[t!]
    \centering
    \begin{subfigure}[t]{0.5\textwidth}
        \centering
        \includegraphics[width = 0.8\textwidth]{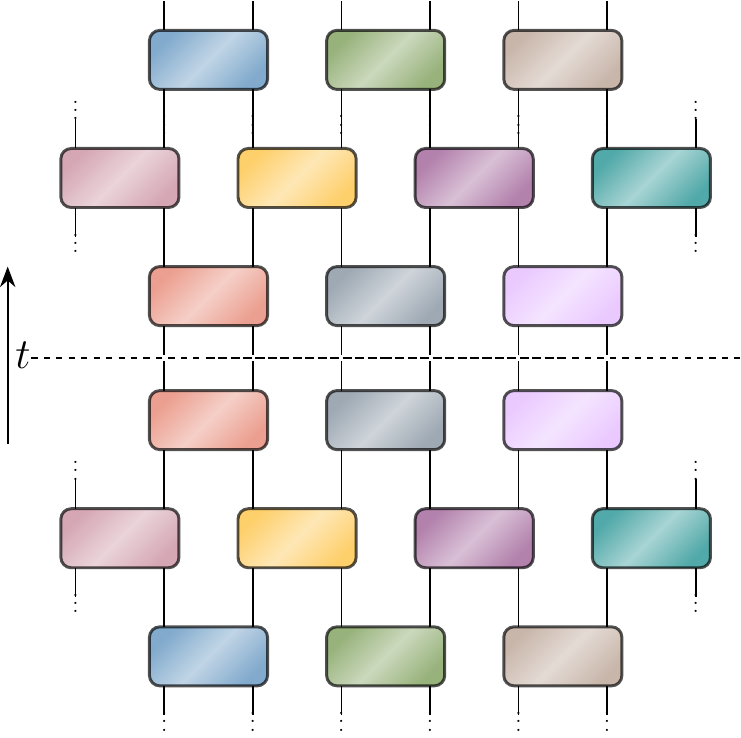}
        \caption{}
        \label{unfoldedglobal}
    \end{subfigure}%
    \begin{subfigure}[t]{0.5\textwidth}
        \centering
        \includegraphics[width = 0.8\textwidth]{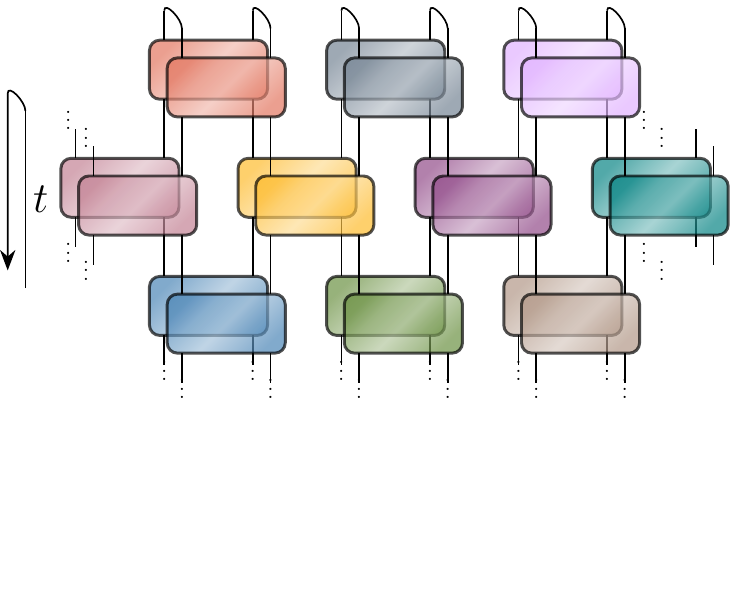}
        \caption{}
        \label{foldedglobal}
    \end{subfigure}
    \captionsetup{justification=Justified, singlelinecheck=false, font=small}
    \caption{\textbf{Circuit representing Global and Local TR Evolution:} The color of the gates indicates whether they are identical or not. Local TR dynamics is implemented by picking the gates to be of the form \( U = V^T V \). The global TR dynamics is implemented by making the full circuit have the structure \( V^T V \), as indicated by the mirrored color of the gates. (a) Circuit laid out in a sheet geometry as indicated by the flow of time. (b) The same circuit visualised in a folded-geometry as shown by the flow of time. This provides a clear approach to average over non-locally spaced gates.}
    \label{globaltr}
\end{figure*}

For concreteness, let us begin by focusing on continuous time dynamics generated by a Hamiltonian $H$.
Without measurements, the unitary time evolution $\ket{\psi(t)} = {\rm e}^{- i H t} \ket{\psi_0}$ is TR-invariant if and only if the time reversal operator $\Tau$ applied to the backwards time-evolved state $\ket{\psi(-t)} = {\rm e}^{+ i H t} \ket{\psi_0}$ is equal to the 
state resulting from forward-evolving the state
$\Tau \ket{\psi_0}$ by a time $t$, 
${\cal T} {\rm e}^{+ i H t} \ket{\psi_0}=$
${\rm e}^{- i H t} {\cal T}\ket{\psi_0}$
(e.g., Ref.~\cite{Sakurai_Napolitano_2020}).
Since this is true for all kets $\ket{\psi_0}$ we obtain
$\Tau H \Tau^{-1} = H$.
To process this condition further, we recall that for any fixed orthonormal basis of Hilbert space, the anti-unitary operator $\Tau$ can be written 
in the form
$\Tau =\hat{U}_0 \hat{K}$
where $\hat{K}$ is the complex conjugation operation (which complex-conjugates the expansion coefficients of each ket in the given basis)
and 
$\hat{U}_0$ is a fixed, but basis dependent unitary operator.
Note that while  both 
$\hat{U}_0$
and $\hat{K}$ are basis dependent, 
$\Tau =\hat{U}_0 \hat{K}$
is not. In particular, in our case of interest where $\Tau^2 = + 1$, there exists~\cite{MethaBook1967} a TR-invariant orthonormal basis 
$\lbrace \ket{i} \rbrace_i$
where $\hat{U}_0 = \mathrm{1}$, and thus $\Tau \ket{i} = \ket{i}$.
In this basis, the TR-invariance condition
thus simplifies to
$\Tau H \Tau^{-1} =$ ${\hat K} H {\hat K}^{-1}=$
$H$, implying
that the matrix $H_{ij} = \braket{i | H}{ j }$ is real and symmetric. In turn, this implies that the unitary evolution operator $U(t) = {\rm e}^{- i H t}$ in this basis is a unitary (in general complex) and symmetric matrix $U_{ij}=$ $\braket{i | U(t)}{ j }$.
One can prove~~\cite{MethaBook1967} that any unitary 
(in general complex)
symmetric matrix is of the form $V^T V$ with $V$ being some unitary matrix. This representation shows that the set of unitary symmetric matrices coincides with the Cartan class AI coset space $U(N)/O(N)$ with $N$ the dimension of the Hilbert space, as $V^T V$ is manifestly invariant under $V \to O V$ for any orthogonal matrix $O \in O(N)$ \cite{CartanSymmetricSpaces,
helgason1978differential, zirnbauer1996riemannian, PhysRevB.55.1142,
PhysRevB.78.195125,Ryu2010Topological}. The set of such matrices is called the Circular Orthogonal Ensemble (COE).

Next, we Trotterize the time evolution generated by a (generic~\footnote{ i.e., ``non-integrable''}) Hamiltonian $H$, discussed in the preceding paragraph,
as sketched in  Fig.~\ref{fig:enter-label}. It is straightforward to show that each gate is unitary symmetric, and thus belongs to the COE. (The gates are all the same.) Upon adding 
weak measurements interspersed between these symmetric gates, this setup defines the dynamics of a continuously monitored TR-invariant system. These measurements are performed in the TR-invariant basis, and are thus on average TR-invariant. Nevertheless, note that such measurements still break the global TR invariance of the evolution, unless the measurement outcomes obey a mirror symmetry in time -- we will come back to this later. Even so, the TR-invariance of the underlying unitary evolution still manifests itself by having the gates be 
represented by unitary symmetric matrices in the TR-invariant basis.

To make analytic progress, we then move away from this Hamiltonian setting, and make the gates random in space and and time, drawn from the COE. This additional randomness can be motivated by considering that the 
measurement outcomes are random anyway, so further making the gates 
random
does not break any additional symmetries. However, it is important to note that even though each local gate
is independently symmetric and thus TR-invariant, the global evolution operator of the resulting quantum  circuit
{\em is not} TR-invariant even in the absence of measurements. We will denote this setup as ``locally TR-invariant'', 
referring to the fact that each local gate by itself implements a TR-invariant time evolution. In order for the global time evolution to be TR-invariant (``global TR invariance'')
even in the absence of measurements, we further need the gates to obey a mirror symmetry in time, as illustrated in Fig.~\ref{globaltr}. In that case, the total time evolution operator 
is~\footnote{in the TR-invariant basis}
of the form $U=U_1 U_2 \dots U_k U_k \dots U_2 U_1$, where 
all $U_i$
are (complex) unitary symmetric matrices,
which can be rewritten as $U=V^T V$ with $V=U_k \dots U_2 U_1$.  In the rest of this paper, we will derive statistical mechanics mappings capturing the dynamics of quantum information in quantum circuits that are either locally, or both globally and locally TR-invariant~\footnote{Note that we will not consider the case of ``global but not local'' TR-invariance, where the total time evolution operator $U=V^T V$ is symmetric, but local gates are not. Throughout this paper, we have in mind a setting where local gates are generated by a TR-invariant Hamiltonian as in Fig.~\ref{fig:enter-label}, and are thus symmetric. }.  

We close this section by emphasizing again that, in the absence of measurements, the globally TR-invariant random unitary circuit introduced above represents a  random circuit formulation of generic TR-invariant unitary dynamics, in the same sense in which the Haar random unitary circuit~\cite{PhysRevX.7.031016,PhysRevX.8.021014,PhysRevX.8.021013} represents a random circuit formulation of generic unitary dynamics breaking time-reversal symmetry. In the case with time-reversal invariance, the {\it global} TR-invariance condition is essential for this to be the case.

\section{Statistical Mechanics Mappings (No Measurements)}
\label{smmodels}

In this section, we derive
a replica statistical mechanics model for the dynamics of quantum information under local or global 
unitary TR-invariant evolution.
(We discuss measurements in the subsequent Section~\ref{miptsection}.) 
We first discuss  averaging over moments of COE gates and employ this to derive the stat-mech model for the local TR case. We then impose an additional global TR symmetry, corresponding to a mirror symmetry in time, and derive the corresponding stat-mech model in that case as well.

\subsection{Setup}
\label{setupnonmonitored}
We consider a 1D chain of length $L$, qudit of dimension $d$ with the initial state $\rho_0 = \ket{\psi_0}\bra{\psi_0}$ \footnote{We always work in the TR invariant basis defined in section \ref{trsection}}. We subject it to discrete-time dynamics generated by a brickwork random quantum circuit with gates drawn from the COE. We express the gates in the manifestly (TR) symmetric basis $U  = V^T V$, where $V\in U(D)$ with $D:=d^2$ is sampled from the Haar ensemble.\\

Using the vectorized notation $\rho \rightarrow \ket{\rho} \rangle$, 
the evolution of the density matrix $\ket \rho\rangle_{t-1}$ at
time $(t-1)$ to $\ket \rho\rangle_t$ at time $t$ can then be written as $\ket\rho\rangle_t = U_t\otimes U_t^* \ket\rho\rangle_{t-1} = (V^T_tV_t) \otimes (V^T_tV_t)^{*} \ket\rho\rangle_{t-1} $, where $U_t = V^T_t V_t$ is the unitary matrix applied at time step $t$ expressed in the TR invariant basis. Note that in addition to the doubling of $U_t$ due to 
ket
($U_t$) and 
bra
($U_t^*$) components 
of the density matrix evolution, we also here have 
an additional
doubling due to $U_t$ itself being expressed as $V_t^TV_t$.\\

One can study various quantities to probe universal features of such dynamics. A non-linear quantity such as the $n^{\rm{th}}$ Rényi entanglement $S_{n,A}$ entropy of $\rho_A = \tr_{\bar{A}}(\rho)$ is a useful one to demonstrate the stat-mech mapping. It is written as
\begin{equation}
\label{renyientropy}
    S_{n,A} = \frac{1}{1-n}\ln(\tr\rho^n_A),
\end{equation}
More precisely, we are interested in the ensemble average denoted by $\bar{S}_{n,A}$ that can be computed using the replica trick~\cite{PhysRevB.100.134203, PhysRevB.99.174205, PhysRevB.101.104301,PhysRevB.101.104302}:
\begin{equation}
\label{renyiavgreplica}
    \bar{S}_{n,A} = \lim_{k\rightarrow 0} \frac{1}{k(1-n)}\mathbb{E}_{U}[\tr(\rho^{n}_{A})^{\otimes k}], 
\end{equation}
where $\mathbb{E}_U$ is the average over all possible evolutions generated by the COE gates $U$ that are implicit in $\rho$. Here we define the quantity
\begin{align}
\label{partitionfunctionA}
    & Z_{A} := \mathbb{E}_{U}[\Tr(\rho^{\otimes nk}\mathcal{S}_{A,n}^{\otimes k})],
\end{align}
where $\mathcal{S}_{A,n}$ is the SWAP operator that implements the partial trace within region $\bar{A}$ and the matrix multiplication of the $n-$fold replicated density matrices within region $A$ in each of the $k$ replicas. One can formally represent this operation on this $n-$fold replicated space as
\begin{align}
\label{boundarycondition1}
\mathcal{S}_{A,n}  = \prod_{x}&\sum_{\{i\}}\ket{i_{g_x(1)}i_{g_x(2)}\dots i_{g_x(n)}}\bra{i_1i_2\dots i_n},\\
\label{boundarycondition2}
 g_x &= \begin{cases}
        (1 2 3 \dots n), \ x\in A,\\
         \qquad e \qquad\;\;\: x \in \bar{A}.
    \end{cases}
\end{align}
Here, $\ket{i}$ represent the TR invariant basis states in the Hilbert space $\mathcal{H}:= (\mathbb{C}^d)^{\otimes L}$, and we have utilized cycle notation for permutations. ($e$'' denotes the identity permutation.)
Note that we have used the notation $Z_A$ in \eqref{partitionfunctionA} anticipating it to correspond to the partition function of the statistical mechanics model that we shall now describe.

\subsection{COE Averaging and Boltzmann Weights- \\
Local TR symmetry}
The main expression needed to compute the partition function \(Z_{A}\) is \(\mathbb{E}_{U\in \text{COE}}(U^{\otimes N}\otimes U^{*\otimes N})\). This problem of averaging over moments of gates sampled from the COE has been previously solved in Ref.~\cite{matsumotogeneral}. In this work, we recast the averaging into a more intuitive graphical form, while still utilizing analytic results therein.

A natural approach to this problem is to reduce the problem of averaging over COE gates to the standard Haar average  
of the unitary group. 
To achieve this, we treat the gate \(V^TV\) as two copies of the unitary gate \(V\) with a pair of legs contracted. Graphically,
\begin{figure}[H]
    \centering
    \includegraphics{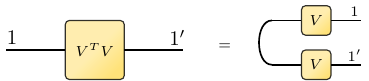},
    \captionsetup{justification=Justified, singlelinecheck=false, font=small}
    \caption{Input and output legs of a given gate labeled by primed and unprimed numbers, respectively.}
    \label{inlinecoecalculus}
\end{figure}
\noindent with primed numbers for the input legs and unprimed numbers for the output legs. Since there are two sources of doubling---one from the bra/ket and another from expressing gates as a product of \( V \) and \( V^T \)---we use different colors for the gates in the bra and ket to make the computations clear.  We also sometimes color the links with the color of the gate they originate from wherever necessary (see for example equation \eqref{eq:inlinecoecalculus2}). The above trick makes it clear that we can treat the problem of averaging over \(U^{\otimes N}\otimes U^{*\otimes N}\) as averaging over twice
the number of copies of the Haar random gate \(V\), with the condition that the legs 
of the two copies of $V$ stemming from the same symmetric unitary $U$ be contracted
(in both, the ket and the bra), as shown in the figure above and 
also in equation \eqref{eq:inlinecoecalculus1}. The averaging is best illustrated graphically as follows  (here $N=2$):
\begin{equation}
\begin{array}{c}
\scalebox{1}{\includegraphics{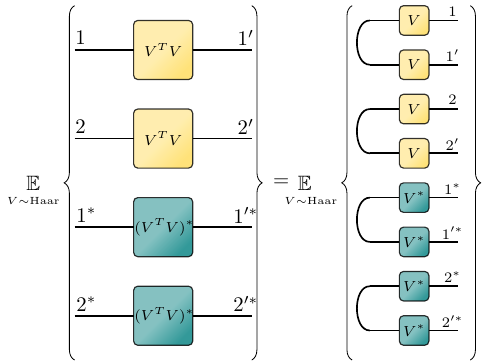}}
\end{array}
\label{eq:inlinecoecalculus1}
\end{equation}
\begin{equation}
\begin{array}{c}
\scalebox{1}{\includegraphics{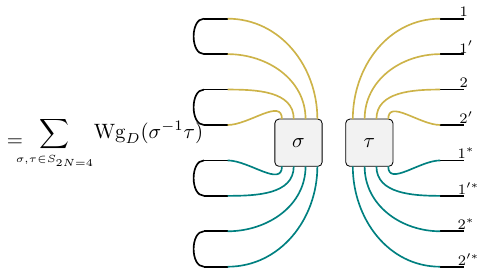}}
\end{array}
\label{eq:inlinecoecalculus2}
\end{equation}
\begin{equation}
\begin{array}{c}
\scalebox{1}{\includegraphics{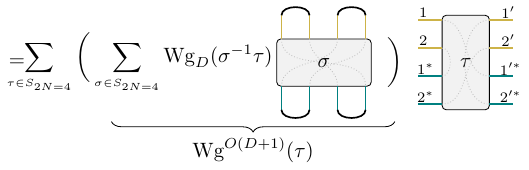}}
\end{array}
\label{eq:inlinecoecalculus3}
\end{equation}

\begin{figure*}
    \centering
    \includegraphics[width=\textwidth]{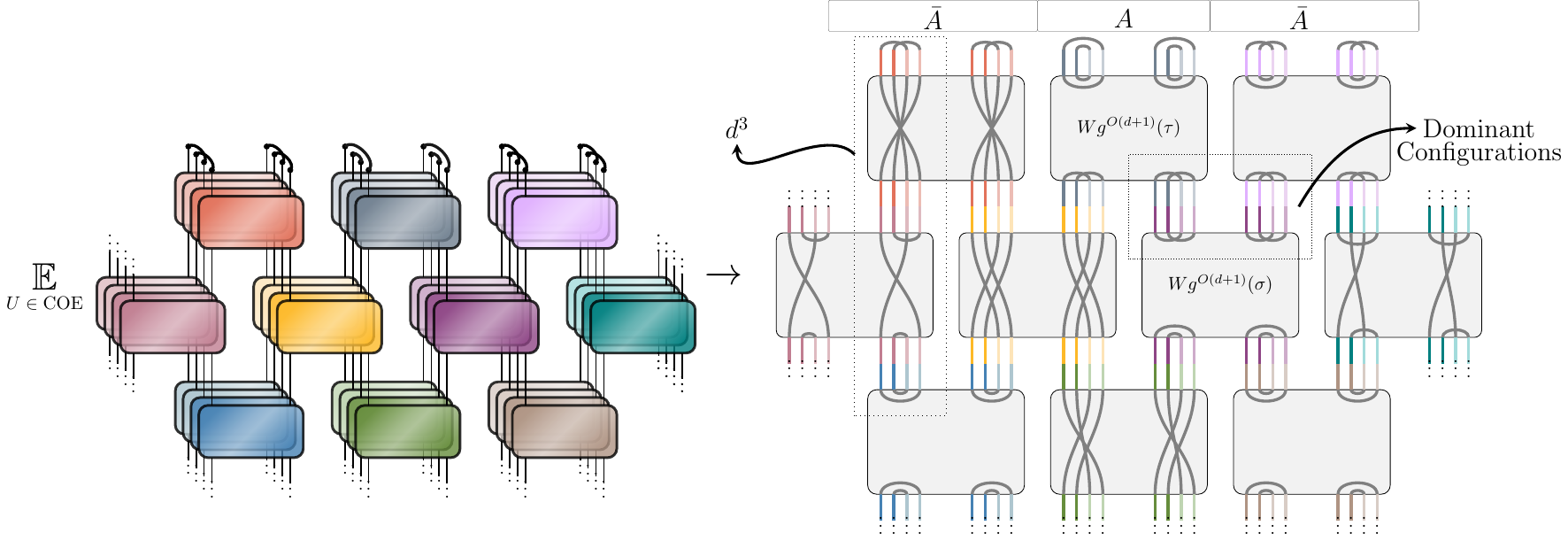}

    \captionsetup{justification=Justified, singlelinecheck=false, font=small}
    \caption{\noindent \textbf{Stat-mech model for Local TR Dynamics. }Left: Replica circuits for Local TR symmetric dynamics with $N=2$. The numbering of the replicas from front to back is $(1,2, 1^*, 2^*)$. Each of the gates within a replica are randomly sampled from the COE and are of the form \( U = V^T V \). We further emphasize the randomness using different colors for non-identical gates and vice-versa. The conjugate replicas are colored with a lighter shade of the color corresponding to the unconjugated ones. The contraction at the top-center depicts region \( A \) where the boundary conditions are \( g_{\rm{SWAP}} \). The contractions at the ends on top (\( \bar{A} \)) are represented with contractions \( e \in S_N \). \noindent Right: A particular configuration of the statistical mechanics model derived from the averaging of the circuit on the left. The links are represented by the color of the respective gates on the left. The links of the conjugated replicas are emphasized using a lighter shade of the corresponding un-conjugated replica links. The grey links pair conjugated links with the unconjugated ones based on the ``spin" \( \sigma \in S_{2N} \) at a given site. Link weights form loops which contribute as $d^{\#\rm{loops}}$ to the partition function while the vertex weights are given by $\mathrm{Wg}^{O(d+1)}(\tau), \; \tau \in S_{2N}$.}
    \label{fullcircavg}
\end{figure*}
\noindent Above, the well-known expression for (unitary) Haar averages was used which brings in the Weingarten function
 ${\rm Wg}_D(\tau\sigma^{-1})$ for the unitary group $U(D)$,
 \begin{eqnarray}
\nonumber
&&\int_{\rm Haar} d \mu(V) \ 
V_{i_1, j_1}
...
V_{i_Q, j_Q}
V^*_{i'_1, j'_1}...V^*_{i'_Q, j'_Q}=
\\  \nonumber
&&=\sum_{\sigma, \tau\in S_Q}
{\rm Wg}_D(\tau\sigma^{-1}) \ 
\delta_{i^{}_1, i'_{\sigma(1)}}
...
\delta_{i^{}_Q, i'_{\sigma(Q)}}
\delta_{j^{}_1, j'_{\tau(1)}}
...
\delta_{j^{}_Q, j'_{\tau(Q)}},
\\ 
\label{LabelEqWeingartenFunctionInComponents}
\end{eqnarray}
where $d \mu(V)$ denotes the unitary Haar measure.

A few comments are in order. The final result \eqref{eq:inlinecoecalculus3} contains Weingarten functions $\text{Wg}^{O(D)}(\cdot)$ for the orthogonal group $O(D)$, first introduced in the context of averaging over moments of orthogonal matrices in \cite{collins_integration_2006}  (see \cite{collins_properties_2009} for a good review on these functions). Interestingly, the orthogonal Weingarten function $\text{Wg}^{O(D+1)}(\cdot)$ here is obtained by summing over a linear combination of the unitary Weingarten functions $\text{Wg}_D(\cdot)$. Although intriguing, this should not be surprising since the COE can be identified with the compact symmetric space $U(D)/O(D)$. However, the presence of $D+1$ instead of $D$ is indeed quite odd (as also noted by \cite{matsumotogeneral}) yet unimportant in our subsequent analysis. An exact Fourier-type expansion for $Wg^{O(D+1)}(\tau)$ is given by \cite{collins_properties_2009}
\begin{equation}
    \text{Wg}^{O(D)}(\tau) = \frac{2^{N}N!}{(2N)!}\sum_{\lambda\vdash N} \frac{f^{2\lambda}}{C'_{\lambda}(D)}\omega^\lambda(\tau),
\end{equation}
where the sum is over integer partitions $\lambda$ of $N$, such that $\lambda = (\lambda_1,\lambda_2,\dots)$ with $\lambda_1\geq\lambda_2\geq\dots\lambda_l\in \mathbb{N}$, $\omega^{\lambda}(\tau)$ are zonal spherical functions \footnote{Zonal spherical functions $\omega^{\lambda}(\tau)$ can be expressed as a linear combination of the characters $\chi^{2\lambda}(\sigma)$ of $\sigma\in S_{2Q}$ associated with the irrep. $\lambda$}, $f^{2\lambda}$ is the dimension of the irreducible representation associated with $\lambda$, and $C'(\lambda)(D) = \prod_{i=1}^{l(\lambda)}\prod_{j=1}^{\lambda_i}(D+2j-i-1)$. \\

Probing equation \(\eqref{eq:inlinecoecalculus3}\) further, we observe a crucial distinction from the case of averaging over (unitary)
Haar gates. Averaging over (unitary) Haar gates results in pairing the states in the input (output) legs of replicas with the input (output) legs of the conjugate replicas. However, averaging over the COE results in mixing between the input and output legs of conjugate replicas. In other words, the input legs of a replica can be paired with the outgoing legs of a conjugated replica. As a result of this, each leg of the ket can be paired with \(2N\) legs of the bra, leading to a sum over \(\tau \in S_{2N}\) in \eqref{eq:inlinecoecalculus3} instead of \(S_N\) in the standard Haar case.
This also makes intuitive sense since averaging over a smaller set (in this case from $U(D) \rightarrow U(D)/O(D)$) would be expected to lead to a richer structure post-averaging.\\

Applying the result \eqref{eq:inlinecoecalculus3} for all gates in the circuit we can interpret $Z_{A}$ as the partition function of
a statistical mechanics model on a square lattice with the permutation in
$S_{2N}$ (``spins") living on the vertices with vertex weights $Wg^{O(D+1)}(\tau)$ (see figure \ref{fullcircavg}). The link weights on the other hand are calculated exactly like in the Haar case by contracting neighboring links and counting the number of loops. An example is 
shown~\footnote{in the light-dotted rectangular box on the left side of the right panel under the letter 
$\bar{A}$}
in figure \ref{fullcircavg}
where we count three loops resulting in the link weight $d^3$. 
However, a difference here is that the loops could possibly be extended in space due to the input/output mixing mentioned earlier. 
Denoting the total number of loops in a configuration $\{g_v\}$ of permutations $g_v\in S_{2N}$ at the vertices of the square lattice by $l_{\{g_v\}}$,
the bulk partition function for $Z_{A}\equiv Z_{\rm{bulk}}$ can be written as

\begin{equation}
    Z_{\rm{bulk}} = \sum_{\{g_v \in S_{2N}\}} d^{l_{\{g_v\}}} 
\prod_{v\in \text{vertices}}\text{Wg}^{O(D+1)}(g_v).
\end{equation}

Similar to the Haar case, the partition function has negative weights due to the negativity of $Wg^{O(D+1)}(\cdot)$ \cite{collins_properties_2009}. The boundary contributions to $Z_A$ are as per \eqref{boundarycondition1} and \eqref{boundarycondition2}: tracing over $\bar{A}$ fixes the permutations in $\bar A$ to be $e=\text{identity}\in S_{N}\subset S_{2N}$ and performing matrix multiplication of $n$ copies (in each of the $k-$replicas) corresponds to fixing the permutations in region $A$ to be $g_{\rm{SWAP}}\in S_{N}\subset S_{2N}$ \footnote{The notation $\tau \in S_{N}\subset S_{2N}$ denotes $\tau$ as a member of $S_N$ embedded in $S_{2N}$. This embedding is given by grouping the legs $\{m,m'\}$ and permuting the resulting $N$ objects as per $S_N$}.\\   

\subsection{Stat-mech Mapping for Global TR Symmetry}
\label{mirrorp0}
As discussed previously, we are interested in the case where we have global TR symmetry in addition to the local TR symmetry already present. As mentioned earlier, we refer to this model as the global TR symmetric model. Such a model requires that both the complete evolution and the local gates be written in the form $U = V^TV$. This can be implemented by adding a ``mirror” to the existing local TR symmetric evolution as shown in figure \ref{unfoldedglobal}. It will also hereon be useful for us to refer to a mirror transformation at a site as the one that takes us from a site to its mirrored counterpart as per figure \ref{unfoldedglobal}.\\  

\begin{figure*}
    \centering
    \includegraphics[width=\textwidth]{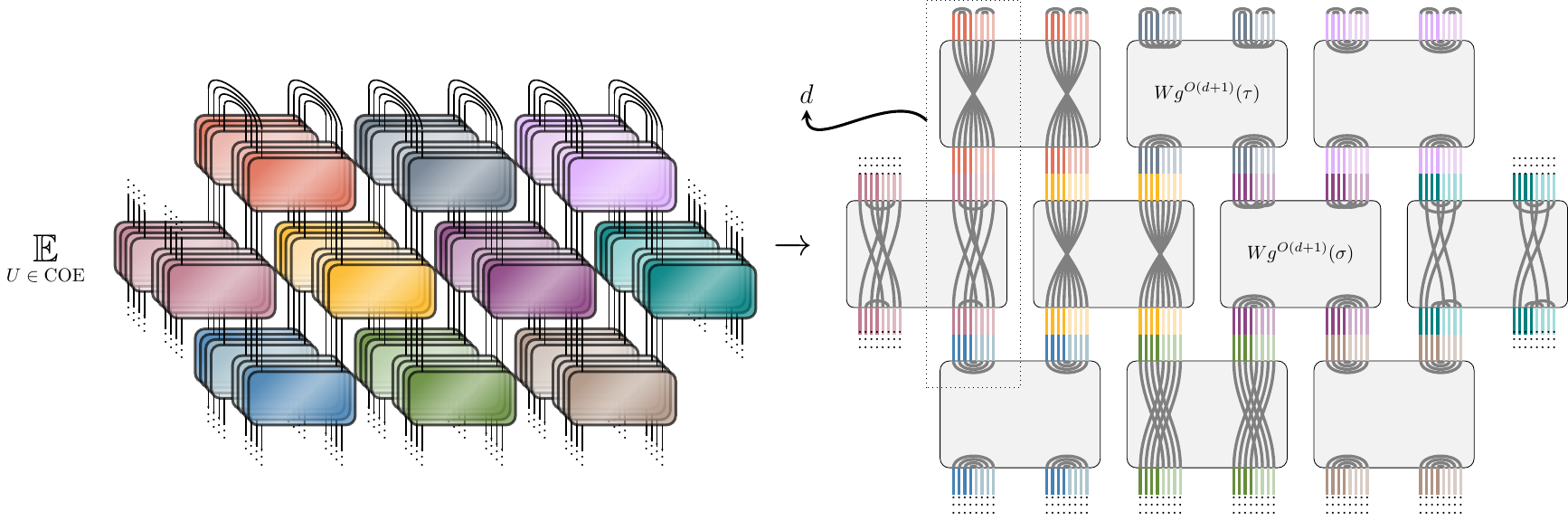}
    \captionsetup{justification=Justified, singlelinecheck=false, font=small}
    \caption{\noindent \textbf{Stat-mech model for Global TR Dynamics. }Left: Replica circuits for Global + Local TR symmetric dynamics with $N=2$. The numbering of the replicas from front to back is $(1,2, 1^*, 2^*, \bar{2}^*, \bar{1}^*, \bar{2}, \bar{1})$, where we represent the mirrored replicas using a bar $(\bar{\cdot})$. Each of the gates within a replica are randomly sampled from the COE and are of the form \( U = V^T V \). We further emphasize the randomness using different colors for non-identical gates and vice-versa. The conjugate replicas are colored with a lighter shade of the color corresponding to the unconjugated ones. The top boundary conditions pair each link with its mirrored counterpart. There are mixed boundary conditions at the bottom as explained in the main text (not shown here). \noindent Right: A particular configuration of the statistical mechanics model derived from the averaging of the circuit on the left. The links are represented by the color of the respective gates on the left. The links of the conjugated replicas are emphasized using a lighter shade of the corresponding un-conjugated replica links. The grey links pair conjugated links with the unconjugated ones based on the ``spin" \( \sigma \in S_{4N} \) at a given site. Link weights form loops which contribute as $d^{\#\rm{loops}}$ to the partition function while the vertex weights are given by $\mathrm{Wg}^{O(d+1)}(\tau), \; \tau \in S_{4N}$.
}
    \label{stat_mech_model_mirror}
\end{figure*}

The stat-mech model requires averaging over non-locally spaced gates, as depicted in Figure \ref{unfoldedglobal}. To address this, we can conceptualize a folded geometry where identical gates occupy the same space-time position (see Figure \ref{foldedglobal}). This simplifies the averaging process: we now have twice the number of replicas of $U$ (or equivalently, twice the replicas of $V$) compared to the local case, and we must perform the average of $U^{\otimes 2N}\otimes U^{*\otimes 2N}$ at each gate location. The resulting expression is identical to \eqref{eq:inlinecoecalculus3}, except we now have 
twice
the number of gates at each site, and the sum runs over $\sigma \in S_{4N}$ instead of $S_{2N}$ \footnote{The $4N$ number of replicas can formally be accounted for as follows: We have 2N replicas as before due to $N$ kets and $N$ bras. In addition, for each of the 2N replicas, we have a mirrored counter-part leading to $4N$ replicas in total.}.
The emergent statistical mechanics model is illustrated in Figure \ref{stat_mech_model_mirror}. Here, the link weights are determined by counting the number of (possibly extended) loops, and the vertex weights are given by $\mathrm{Wg}^{O(D+1)}(\tau)$ with $\tau \in S_{4N}$.\\

The folded geometry begets several peculiarities to the global TR model. In the folded geometry, the transfer matrix of the corresponding statistical mechanics model can be interpreted as a super-operator that acts on a quantum channel and returns a new quantum channel. Indeed, the ``initial state'' of this folded evolution is a quantum channel that maps initial to final state in the original quantum evolution, and the ``seam''
(at the top of Fig.~\ref{globaltr} (b))
at the final step of the folded evolution can be interpreted as the identity quantum channel.

Secondly, we note that each space-time point in this stat-mech model corresponds to two different instances in real-time evolution. Consequently, any operation performed in the stat-mech model will correspond to the same operation applied at two points related by a mirror transformation in the real-time circuit. Furthermore, in the stat-mech model, ``time" can be considered the vertical distance from the bottom or top end of the model in Figure \ref{stat_mech_model_mirror}, which does not correspond to time in the actual circuit dynamics. The implications of these differences become evident when calculating the anisotropy factor, as discussed in Appendix \ref{anisotropycalc}, where we outline a protocol to numerically extract correlation functions in this stat-mech model.\\

Another noteworthy difference from local TR due to the folded geometry are the boundary conditions. Here we have mixed boundary conditions at the bottom because the initial and final states of the real-time evolution correspond to the same location in the stat-mech model. At each gate, half the links have free boundary conditions due to the freedom in the choice of the initial state whereas the other half have their permutation fixed as $g_{\rm{SWAP}} (e)$ in $A(\bar{A})$ respectively. At the seam (top), the boundary conditions are fixed by pairing both bra and ket in each replica with their corresponding mirrored counter-part in accordance with the mirror symmetry (see figure \ref{stat_mech_model_mirror}).\\

\section{Application: MIPT in monitored TR-invariant systems}
\label{miptsection}
Random quantum circuits and their stat-mech mappings have proven invaluable in addressing various dynamical questions in quantum many-body systems \cite{Potter_2022, annurev:/content/journals/10.1146/annurev-conmatphys-031720-030658}. These mappings have been successfully applied to study entanglement growth \cite{PhysRevX.7.031016, PhysRevB.99.174205}, operator spreading hydrodynamics~\cite{PhysRevX.8.021014}, measurement-induced phase transitions (MIPTs) \cite{PhysRevB.101.104301, PhysRevB.101.104302}, error-correction \cite{PRXQuantum.5.020343},  and the complexity of classical computations \cite{PhysRevX.12.021021}. In this work, we explore the application of the aforementioned stat-mech models for the study of TR-symmetric MIPTs. We choose MIPTs as our focus due to their versatility in applications and interpretations \cite{PhysRevLett.129.200602, PhysRevLett.125.030505, PRXQuantum.5.020304, PhysRevX.10.041020}, making them an ideal starting point for demonstrating the utility of our stat-mech models. Like in the previous section, we give an overview of how the previously derived stat-mech models are altered once we include measurements for the local TR case followed by the global TR case. We then discuss the symmetry properties of these models and end by discussing some analytic results in the limit of large onsite Hilbert space dimension $d$.

\subsection{MIPT Setup}
The basic setup is identical to the one in section \ref{setupnonmonitored}. We consider a brickwork random quantum circuit with gates being sampled from the COE. Such a circuit represents discrete time dynamics on a qudit spin chain of length $L$ with Hilbert space dimension $d$ at each site. In addition, we include projective measurements at every time-step with a probability $p$ on each qudit. We define a single realization of a circuit to be an instance of randomly picked COE gates and measurement locations. Starting with an initial state $\rho_0 := \ket{\psi_0}\bra{\psi_0}$ one can then write the overall dynamics of a realization as a quantum trajectory $\rho_{\textbf{m}} = K_{\textbf{m}}\rho_0 K_{\textbf{m}}^{\dag}$. Here we use $\mathbf{m}$ for the list of measurement outcomes and $K_{\mathbf{m}}$ for the Kraus operators that contain unitary gates and projective measurements, with $K_{\mathbf{m}}$ satisfying the usual Kraus relation $\sum_{\mathbf{m}}K_{\mathbf{m}}^{\dag}K_{\mathbf{m}} = \mathbb{I}$.\\

Our goal is to study the dynamics of the above setup. In particular, we probe the universal features of the dynamics using the $n^{\text{th}}$ Rényi entanglement $S_{n,A}$ entropy \eqref{renyientropy}. Without the measurements, it is well known that the final steady state of a typical circuit has a volume law with $S_{n,A}\sim L$ in any given subsystem of size $L$. The measurements inhibit this growth of entanglement and there is a critical measurement rate $p_c$ above which the steady state exhibits area-law scaling $S_{n,A}\sim \mathcal{O}(1)$. This is indeed the measurement-induced phase transition. It is crucial to note that this transition takes place in the nature of the quantum trajectories $\rho_{\textbf{m}}$ where the experimentalist keeps track of the measurement outcomes $\mathbf{m}$. The transition is invisible in the measurement averaged density matrix, such as in the case of decoherence. Consequently, we first evaluate properties of a given trajectory $\rho_{\textbf{m}}$ and only then average over the gates, measurement outcomes, and locations.  \\

Following \cite{PhysRevX.12.041002}, we use the notation $\mathbb{E}_{U}(.)$, $\mathbb{E}_{{X}}(.)$, and $\sum_{\mathcal{M}(\mathbf{X})}(.)$ to denote averaging over COE gates, measurement locations, and over the measurement outcomes 
\an{\bf at}
a given set of locations \textbf{X} respectively. With a slight abuse of notation, we will also sometimes combine the average over the set  $\{\mathbf{X}, \mathcal{M}(\mathbf{X})\}$ as $\sum_{\mathbf{m}}p_{\mathbf{m}}$. The average Rényi entanglement entropy we aim to compute can then be written as
\begin{equation}
\label{renyidefinition}
    \bar{S}_{n, A} = \mathbb{E}_{U}\sum_{\mathbf{m}}\bigg\{p_{\mathbf{m}}\times \frac{1}{1-n}\ln\bigg[\frac{\tr(\rho^n_{A,\mathbf{m}})}{\tr(\rho_{\mathbf{m}})^n}\bigg] \bigg\},
\end{equation}
where we weight each trajectory $\rho_{\mathbf{m}}$ with its corresponding Born probability $p_{\mathbf{m}} = \Tr(\rho_{\mathbf{m}})$. The computation of the above 
average can be done using the Replica trick as shown in  \cite{PhysRevB.101.104301, PhysRevB.101.104302, PhysRevB.100.134203, PhysRevB.99.174205, PhysRevX.12.041002} 
using
\eqref{renyiavgreplica} as
\begin{equation}
\label{renyiavgreplica2}
    \bar{S}_{n,A} = \frac{n}{1-n} \lim_{N\rightarrow 0}\mathbb{E}_{X}\bigg[\frac{Z_{A}(\mathbf{X})-Z_{0}(\mathbf{X})}{N}\bigg]
\end{equation}
where,
\begin{align}
\label{partitionfunctionmipt1}
    & Z_{A}(\textbf{X}) = \mathbb{E}_{U}\sum_{\mathcal{M}(\mathbf{X})} [\Tr\{(K_{\textbf{m}}\ket{\psi}\bra{\psi} K_{\textbf{m}}^{\dag})^{\otimes (N+1)}\mathcal{S}_{n,A}^{\otimes k} \}]\\
\label{partitionfunctionmipt2}
    & Z_{0}(\textbf{X}) = \mathbb{E}_{U}\sum_{\mathcal{M}(\mathbf{X})}  [\Tr\{(K_{\textbf{m}}\ket{\psi}\bra{\psi} K_{\textbf{m}}^{\dag})^{\otimes (N+1)}\}],  
\end{align}
similar to equation \eqref{partitionfunctionA}. $Z_{A,0}(\mathbf{X})$ are identical in the bulk and differ in their boundary conditions being $S_{n,A}^k$ and $\mathbb{I}$ respectively. $N:=nk$ is the number of replicas where replicas and the additional replica has entered by absorbing the $p_{\mathbf{m}}$ factor in \eqref{renyidefinition} as $\Tr(\rho_{\mathbf{m}})$. Finally, note that we do not include the average over the measurement locations $\mathbf{X}$ in the above partition functions. This is because the statistical mechanics model for the MIPT will also have non-local weights that depend on $\mathbf{X}$, similar to the case with $p=0$ (no measurements)
discussed in the previous section.

\subsection{Boltzmann Weights}
\label{boltzmannweightsmipt}
\subsubsection{Local TR symmetric Model}
\label{unmirroredmipt}
In order to obtain the statistical mechanics model, we need to evaluate the combination $\mathbb{E}_{U}\sum_{\mathcal{M}(X)}K^{\otimes (N+1)}_{\mathbf{m}}\otimes K^{*\otimes (N+1)}_{\mathbf{m}}$ from \eqref{partitionfunctionmipt1} and \eqref{partitionfunctionmipt2}. This average can be decomposed into separate averages: $\mathbb{E}_{U}$ over the COE gates and $\sum_{\mathcal{M}(X)}$ over the measurement outcomes, which only involve the projection operators $P_i:=\ket{i}\bra{i};\;\;i\in\{1,\dots,d\}$ in the TR invariant basis. The gate averaging $\mathbb{E}_{U}$ has already been discussed in equations \eqref{eq:inlinecoecalculus1}-\eqref{eq:inlinecoecalculus3} and remains unchanged, resulting in the model as shown in figure \ref{fullcircavg}. In addition to this, we have projection operators at the locations $\mathbf{X}$ that freeze the links in all the replicas at all $X\in \mathbf{X}$ to the measurement outcome $\mathcal{M}(X)$, resulting in a factor of 1 in the partition function \footnote{The factor 1 is due to $\tr M^{\dag}M$, where $M$ is the measurement operator that projects onto the state $\ket{\mathcal{M}(X)}$.}. As per the stat-mech model in figure \ref{fullcircavg}, the measurement at point $X$ will consequently fix all the links that are part of the loops containing the links at $X$ to the same outcome $\mathcal{M}(X)$. Therefore any further measurement made at another part of the loops going through $X$, say $X'$, will only contribute when $\mathcal{M}(X)=\mathcal{M}(X')$. Performing the summation $\sum_{\mathcal{M}(X)}$ will then result in a factor of $d$ corresponding to the set of loops that pass through $X$. \\
The statistical mechanics model for the 
TR-invariant MIPT
can then be summarized as follows: we have a square lattice with permutation degrees of freedom $\sigma$ living on the vertices with vertex weights $\text{Wg}^{O(D+1)}(\sigma)$, exactly like in the $p=0$ case. The unmeasured link weights are calculated by contracting neighboring gates and counting the number of loops (assuming every link in the loop is unmeasured), with each loop contributing a factor of $d$. If a measurement is made at a given $X$, all the loops that contain any of the $2(N+1)$ links at $X$ reduce to a total factor of $d$. Further measurements made at any other location containing links that form these frozen loops do not contribute. 
Denoting, for every configuration $\{g_v\}$ of permutations at the vertices of the square lattice, 
the total number of loops that contain all unmeasured 
links by $l_{\{g_v\}}(\mathbf{X})$ and the total number of measurement locations that do not have common loops by $m_{\{g_v\}}(\mathbf{X})$,
the bulk partition function for $Z_{A, 0} \equiv Z_{\rm{bulk}}(\mathbf{X})$ is given by
\begin{multline}
\label{partitionbulkmiptnonmirrored}
        Z_{\mathrm{bulk}}(\mathbf{X}) = \sum_{\{g_v \in S_{2N}\}} 
d^{[l_{\{g_v\}}(\mathbf{X})+m_{\{g_v\}}(\mathbf{X})]} 
\\\prod_{v\in \text{vertices}}\text{Wg}^{O(D+1)}(g_v).
\end{multline}\\
The boundary conditions of $Z_{A}$ are fixed as per \eqref{boundarycondition1} and \eqref{boundarycondition2}, whereas the boundary conditions for $Z_{0}$ are $e\in S_{N+1}\subset S_{2(N+1)}$ everywhere on the boundary.\\
We find a volume law to area law transition for this model at $p_c \approx 0.15$ (see numerical data in figure \ref{tmi} and corresponding section \ref{numerics}). We delay the discussion of the critical properties and analytic results in the large $d$ limit to later sections \ref{universality} and turn to discussing the MIPT stat-mech model for when the global TR symmetry is added.

\subsubsection{Global TR Symmetric Model}
\label{globaltrmipt}

The global TR symmetric $p=0$ dynamics is modeled by the circuit in figure \ref{unfoldedglobal}. As we shift $p$ away from 0, we immediately disrupt the mirror symmetry essential for global TR dynamics, given the random nature of measurements. To preserve global TR, we must therefore implement measurements that respect the mirror symmetry. This means that for every measurement made at a specific space-time point in the first half of the evolution, a corresponding forced measurement must be made at its mirrored counterpart. Furthermore, we must also post-select the outcomes in the mirrored half to match those in the first half of the evolution. This is a stringent requirement and we investigate later on whether it is possible to relax this.

Due to post-selection of measurements, the resultant stat-mech model does not follow directly from \eqref{partitionfunctionmipt1} and \eqref{partitionfunctionmipt2}, since
for
the 
Born probability 
we now have in general $\tr(\rho_{\mathbf{m}})\neq p_{\mathbf{m}}$, whereas there was an equality in the local TR symmetric case.
This is because $\rho_{\mathbf{m}}$ is the state after the complete global TR symmetric evolution whereas $p_{\mathbf{m}}$ is calculated only from the first half of the evolution. Denoting the full evolution as $O_{\mathbf{m}} = K^T_{\mathbf{m}}K_{\mathbf{m}}$ where $K_{\mathbf{m}}$ is the first half of the full-evolution, one can then re-write \eqref{partitionfunctionmipt1} and \eqref{partitionfunctionmipt2} for the global TR symmetric model as 
\begin{align}
&\begin{aligned}
    \label{partitionfunctionmipt1global}
    Z_{A}(\textbf{X}) = \mathbb{E}_{U}\sum_{\mathcal{M}(\mathbf{X})} \big[\Tr(O_{\textbf{m}}\ket{\psi}\bra{\psi} O_{\textbf{m}}^{\dag})^{\otimes N}\mathcal{S}_{n,A}^{\otimes k}\\\times \Tr(K_{\textbf{m}}\ket{\psi}\bra{\psi} K_{\textbf{m}}^{\dag}) \big] 
\end{aligned} \\
&\begin{aligned}
    \label{partitionfunctionmipt2global}
    Z_{0}(\textbf{X}) = \mathbb{E}_{U}\sum_{\mathcal{M}(\mathbf{X})} \big[\Tr(O_{\textbf{m}}\ket{\psi}\bra{\psi} O_{\textbf{m}}^{\dag})^{\otimes N}\\\times \Tr(K_{\textbf{m}}\ket{\psi}\bra{\psi} K_{\textbf{m}}^{\dag}) \big],
\end{aligned}
\end{align}
where $p_{\mathbf{m}} = \Tr(K_{\textbf{m}}\ket{\psi}\bra{\psi} K_{\textbf{m}}^{\dag})$. The resultant stat-mech model can be developed step-by-step as follows: for $p=0$, the stat-mech model is the one in figure \ref{stat_mech_model_mirror} with $2N$ replicas. 
For $p\neq 0$ we need to make two changes to this model. Firstly, we get an additional replica due to $p_{\mathbf{m}}$ resulting in a total of $2N+1$ replicas \footnote{Note that the reason for the Born replica not being doubled is that the trajectories are fixed (due to post-selection of the measurement outcomes) in the second half of the evolution.}. Secondly, given a set of measurement locations $\mathbf{X}$, one must reduce all the loops that cross the $2(2N+1)=4N+2$ links in $X\in\mathbf{X}$ to a factor of $d$ (instead of just $2N+1$ as in the local model). The reasoning for this is the same as for the local TR MIPT model discussed previously. Measurements freeze the links at $X$ (and now also its mirrored counterpart) in all the replicas to the same measurement outcome $\mathcal{M}(X)$, resulting in weight $1$ in the partition function. We then sum over $d$ possible measurement outcomes resulting in the total weight $d$ in the partition function. Additionally, if there is a measurement at some other location $X'$ involving the loops that cross through $X$, they only contribute when $\mathcal{M}(X)=\mathcal{M}(X')$. Finally, out of the $2N+1$ total replicas, $2N$ replicas have the same boundary conditions as those for the model at $p=0$: mixed boundary conditions at the bottom with half the legs being free and the other half being fixed as $g_{\rm SWAP}$ and $e$ in regions $A$ and $\bar{A}$ respectively, while the boundary conditions at the ``seam" (on the top) are fixed by pairing both the bra and ket in each replica with their mirrored counterparts. The last Born-probability replica has $e$ boundary conditions at the top and free boundary conditions at the bottom. \\
We find that this model has an entanglement transition at a critical measurement probability of $p_c \approx 0.15$ as well (see numerical data in figure \ref{tmi} and corresponding section \ref{numerics}). However, as we shall now discuss, the two models have different symmetry properties and fall into different universality classes.

 We summarize and close this section by highlighting
the folded formulation as the natural description of 
a TR-invariant circuit. As will be emphasized later in this paper, the global TR invariant circuit is the only version of the circuit that in fact possesses  TR invariance, and this version has a folded description. Therefore, by default, a (global) TR-invariant circuit can always be described in  the folded formulation, and we may consider this the defining property of a TR-invariant circuit.
As mentioned (and as will be stressed again later), a global (or ``strong'') version of TR-invariance enforces the presence of
mirror symmetry in each quantum trajectory, and therefore the first half of the circuit evolution is the only independent evolution, completely determining the second half of the evolution. This is 
expressed in
(\ref{partitionfunctionmipt1global}) and (\ref{partitionfunctionmipt2global})
by 
writing the full evolution as  $O_{\mathbf{m}} = K^T_{\mathbf{m}}K_{\mathbf{m}}$. As visualized in Figs.~\ref{globaltr}, \ref{stat_mech_model_mirror}, we can think of the evolution of the 
(global) TR-invariant circuit as acting on a doubled Hilbert space for each, the ``ket'' as well as the ``bra'' evolution; the doubling corresponds to the tensor product of Hilbert spaces associated with the first and second half of the evolution (i.e. of the two mirrored halves). Written out in terms of the vectorization~\footnote{Where
$\langle\langle j, j^* | i, i^* \rangle\rangle = \delta_{i, j}\delta_{i^*, j^*}$} of the
density matrix $\rho_t =|\psi_t\rangle \ \langle \psi_t|$ $\leftrightarrow$
$\ket{\rho_t}\rangle=$
$\sum_{i, i^*}\psi_i(t) \  \psi^*_{i^*}(t) \ \ket{i, i^*}\rangle$, Sec.~\ref{setupnonmonitored} (2nd paragraph),
Fig.~\ref{stat_mech_model_mirror}
reads in the presence of measurements ($t\geq 0$)
\begin{eqnarray}
\nonumber
&&
\langle \langle {{\rm 
\scriptstyle 
Seam}} |
\Bigl [ (K_{\mathbf{m}}\otimes K_{\mathbf{m}}^*) \otimes 
(K_{\mathbf{m}}\otimes K_{\mathbf{m}}^*)\Bigr] \bigl (|\rho_{-t}\rangle\rangle\otimes |j, j^*\rangle\rangle \bigr)=
\\ 
\nonumber
&&
=
\langle\langle j, j^*| \ 
\Bigl [ (K^T_{\mathbf{m}} K_{\mathbf{m}}) \otimes 
(K^{*T}_{\mathbf{m}}K_{\mathbf{m}}^*)\Bigr]
\ 
|\rho_{-t}\rangle\rangle
=
\langle\langle j, j^*| \rho_{+t}\rangle\rangle,
\\ 
\label{LabelEqMirrorDoubling}
&&
\end{eqnarray}
\noindent where the seam corresponds to the state
\begin{eqnarray}
\label{LabelEqSeamState}
&&\langle \langle {\rm {\scriptstyle Seam}} | := \sum_{k, k^*}
\langle\langle i, k^*| \otimes \langle\langle i, k^*|.
\end{eqnarray}
The action in the first line of (\ref{LabelEqMirrorDoubling})
on the seam
(\ref{LabelEqSeamState}) as the ``initial state''
from the left with the evolution operator
$\Bigl [ (K_{\mathbf{m}}\otimes K_{\mathbf{m}}^*) \otimes 
(K_{\mathbf{m}}\otimes K_{\mathbf{m}}^*)\Bigr]$
 of the folded circuit  can be viewed as providing  a definition of the 
 (global) TR-invariant circuit on the doubled Hilbert space by an amount of time 
$\boldsymbol{\mathfrak{t}}=t \geq 0$ (half the total time $2t$). It has a ``built-in'' $Z_2 \times Z_2$ symmetry in 
every quantum trajectory,  exchanging the two mirrored halves of the folded circuit independently in the evolution of the ``ket'' 
($K_{\mathbf{m}}\otimes K_{\mathbf{m}}$)
and of  the ``bra''
($K_{\mathbf{m}}^*
\otimes K_{\mathbf{m}}^*)$, which represents the ``built-in'' TR symmetry of the folded formulation.
[As will be described in Sec.~\ref{globaltrmipt2}
below, this will lead to a characteristic  $H_N\times H_N$
symmetry
of the statistical mechanics model of the (global)  TR-invariant circuit.]
For the so-described circuit, Eq.~\ref{LabelEqMirrorDoubling} expresses
the statements in the  last three paragraphs of Sec.~\ref{mirrorp0},  when measurements are present. 
Clearly, in the folded formulation, the special post-selection requirement for the measurement outcomes in the second (mirrored) half of the evolution to match precisely those of the first half (as mentioned in the first paragraph of Sec.~\ref{globaltrmipt}), is simply replaced by working with the second tensor power 
$(K_{\mathbf{m}}\otimes K_{\mathbf{m}}^*) \otimes 
(K_{\mathbf{m}}\otimes K_{\mathbf{m}}^*)$
of the evolution operator of the first half of the circuit,
$(K_{\mathbf{m}}\otimes K_{\mathbf{m}}^*)$. Since the calculation of 
averages of tensor powers of the circuit evolution operator
(of the type mentioned, e.g., in the first sentence of Sec.~\ref{unmirroredmipt})
is already a necessary ingredient for formulating observables relevant for the circuit dynamics, the additional doubling of the tensor power in the folded circuit corresponds, conceptually, to no additional novel feature or ingredient, or ``post-selection'' effort.  
(But this doubling has impact on symmetry of the statistical mechanics model, as mentioned and  discussed in Sec.~\ref{globaltrmipt2}).\\
- Finally, we note that the folded formulation of the (global) TR circuit makes the emergence of conformal symmetry at the transition, that requires the ability to exchange the roles of circuit space and time, more understandable, which would be more difficult to formulate in the post-selected, not-folded description of the same circuit. Similarly, the folded formulation also exhibits a link between global TR-invariant circuits and circuits possessing spatial reflection symmetry $x \to -x$ in every quantum trajectory.

\subsection{Universality and Symmetry}
\label{universality}
\subsubsection{Local TR Model}
\label{LabelSubSubSectionLocalTRModel}
The symmetry groups of stat-mech models corresponding to MIPTs dictate their critical properties. In this section, we examine the symmetry properties of both the MIPT stat-mech models with TR symmetries  discussed in section ~\ref{boltzmannweightsmipt}. In the following, we use lower case $n$ when referring to groups more broadly (with no direct reference to MIPT or the stat-mech model in general) and reserve upper-case $N$ for the number of replicas.

A key group that emerges in the study of these models' symmetries is the hyperoctahedral group $H_{n}$, a sub-group of $S_{2n}$. Formally, $H_n$ is introduced by considering the labeling
of $2n$ objects by
$\{1,1',2,2',...,n,n'\}$. 
(Here we use a notation as in Fig.~\ref{eq:inlinecoecalculus1},
where $l$ and $l'$ denote, e.g.,  the output and the input leg of the gate in replica number $l$.)
$H_n$ is then defined as the set of 
\begin{enumerate}
    \item Single transpositions $(l \; l')$ for $1 \leq l \leq n$
    \item Double transpositions $(l \; m ) (l' \; m')$ for $1 \leq l,  m  \leq n$.
\end{enumerate}

\noindent 
In our problem, $H_N$
can be understood as being generated by the operations that permute replicas (giving the $S_N\subset H_N$ subgroup of $H_N$) and operations that permute the input/output legs of gates within each 
replica~\footnote{The cardinality of $H_N$ can thus be verified to be $2^N N!$. There are $N!$ permutations of the group $S_N$, and each permutation has 2 internal permutations from switching the legs of the symmetric gate, resulting in 
$2^N N!$ operations.}.\\

These operations indeed keep eqs.~\eqref{eq:inlinecoecalculus1}-\eqref{eq:inlinecoecalculus3} invariant, even before averaging step \eqref{eq:inlinecoecalculus2}. 
$H_N$ generators
can also be seen graphically as being implemented by 
\begin{figure}[H]
    \centering
    \label{hnfigurelabelling}\includegraphics{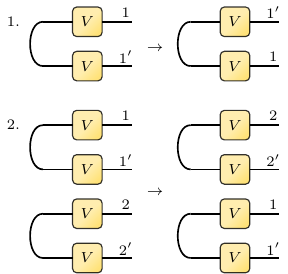}.
    \label{inlinecoecalculus}
\end{figure}
We emphasize that while the latter (double transposition) is the standard replica permutation symmetry that exists even for the unsymmetric Haar case, the former (single transposition) is a manifestation of the symmetric structure imposed by time-reversal. \\
It turns out that the 
vertex weights $\text{Wg}^{O(D+1)}(\tau)\;(\tau\in S_{2N})$ are also 
symmetric under the action of the 
group $H_N \times H_N$ \cite{matsumotogeneral} \footnote{
\label{wgproperty} 
Mathematically, this can be seen as follows. $\text{Wg}^{O(D+1)}(\tau)$ comprises of a linear combination of the unitary Weingarten function $\text{Wg}_D(\sigma^{-1}\tau)$ and the graphical contraction being summed over in equation \eqref{eq:inlinecoecalculus3}. The unitary Weingarten functions are symmetric under $S_{2N}\times S_{2N}$ since they are class functions. The contractions, on the other hand, define the \textit{coset type} of the double cosets of the group $H_N\times H_N$ in $S_{2N}$ \cite{matsumotogeneral} (see Appendix \ref{H_Qgroup}). Since coset types of two permutations $\sigma,\tau\in S_{2N}$ coincide iff they are connected by the conjugate action of $H_N\times H_N$ by definition, such that $\sigma = h_1\tau h_{2}^{-1}, \; h_1,h_2\in H_{N}$ , the graphical contractions are symmetric under $H_{N}\times H_{N}$. Consequently, $Wg^{O(D+1)}(\tau)$ is constant over the different double cosets of $H_N\subset S_{2N}$. }. An intuitive way to see this is to first note  that the contraction block in \eqref{eq:inlinecoecalculus3} counts the number of loops. There exist two operations that keep the number of loops in this block constant. These are: 
\begin{enumerate}
    \item Exchanging any two consecutive set of legs that are connected by the external black loops originating from the forced contraction of $U^T$ and $U$. 
    \item Transformations 
    $\sigma \in S_{N}\subset S_{2N}$
    that permute the pairs of these consecutive legs between the ket (yellow) and the bra (blue) side.
\end{enumerate}
The above transformations can be directly compared to the two generators 
of $H_N$ outlined earlier, leading us to conclude that the contractions 
and therefore $\text{Wg}^{O(D+1)}(\tau)$ are symmetric under $H_N \times H_N$.
\\

Consequently, in the local TR MIPT model, the Weingarten functions are symmetric under $H_{N+1}\times H_{N+1}$, where we recall that the additional replica is due to the Born probability factor.  While in Clifford and Haar ensemble, the link weights obey the symmetry of the respective Weingarten functions \cite{gross_schurweyl_2021, PhysRevB.109.174307}, this does not hold true for the link weights of the local TR model as we shall now explain. We first note that the link weights are at minimum symmetric under $S_{N+1} \times S_{N+1}$ since this symmetry exists microscopically for each circuit and originates from invariance under exchanging the replica copies of the circuit in the bra and ket respectively. This symmetry is also trivially respected by measurements, as they contribute in the form of scalars to the partition function and admit no change when exchanging replicas. Since we know that the vertex weights (ie the Weingarten functions) are symmetric under $H_{N+1}\times H_{N+1}$, we need to only check whether the link weights are symmetric under permutations that are not in $S_{N+1}\times S_{N+1}$ but in $H_{N+1}\times H_{N+1}$ in order to obtain the symmetry group of the full model. The permutations of this type are generated by the transpositions
$(l\;l'),\;1\leq l\leq (N+1)$ 
listed earlier, which exchange the input and output legs at each gate in both the bra and the ket. In figure \ref{fullcircavg}, these types of transformations generically involve ``tearing" and re-joining loops, which does not generally preserve the number of loops.  Consequently, the symmetry group for the local TR MIPT model is $S_{N+1}\times S_{N+1}$, placing it in the same universality class as that of the Haar MIPT.\\ 

One might have intuitively expected a larger symmetry group for ensemble-averaged local TR symmetric dynamics, given the richer structure emerging after averaging as suggested by equation \eqref{eq:inlinecoecalculus3}. An example of this phenomenon is the Clifford ensemble, where the symmetry group of the MIPT stat-mech model is the orthogonal stochastic group $O_{N+1}(p) \times O_{N+1}(p)$, which contains the subgroup $S_{N+1} \times S_{N+1}$, with $d=p^L$, where $p$ is prime, and $L$ represents the length of the qudit chain~\cite{gross_schurweyl_2021, PhysRevB.109.174307}. 
In fact, for local TR dynamics, we do indeed have an $H_{N+1} \times H_{N+1} \supset S_{N+1} \times S_{N+1}$ symmetry at the level of gate averaging. This occurs because exchanging the input and output legs
($(l\;l'), \;\;1 \leq l \leq (N+1)$) 
leaves equation \eqref{eq:inlinecoecalculus1} invariant. However, this symmetry is lost in the statistical mechanics model once the averaging is applied globally to the entire circuit, reducing the symmetry back to $S_{N+1} \times S_{N+1}$.\\

In summary, the hyperoctahedral group $H_n$ is a symmetry of the $n$-copy COE gate averaging \eqref{eq:inlinecoecalculus1}-\eqref{eq:inlinecoecalculus3} and associated with the TR symmetry in the quantum problem. However, when we evaluate the stat-mech model for the local TR MIPT using results from COE gate averaging, the symmetry is reduced to the permutation subgroup $S_{N+1}\subset H_{N+1}$. This results in an overall symmetry group of $S_{N+1}\times S_{N+1}$ for the local TR model — just like in the non-TR invariant Haar case.

\subsubsection{Global TR Model(s)}
\label{globaltrmipt2}

Recall that the major change in going from local to the global TR model is to change the number of replicas from $N+1$ to $2N+1$. Consequently, the stat-mech model has $S_{2N+1}\times S_{2N+1}$ symmetry generated by exchanging any two replica copies \footnote{Similar to the local case, the Weingarten functions $\mathrm{Wg}^{O(D+1)}(\tau)$ are now symmetric under $H_{2N+1}\times H_{2N+1}$ but the link weights limit the symmetry to $S_{2N+1}\times S_{2N+1}\subset H_{2N+1}\times H_{2N+1}$.}. Measurements also respect this symmetry since they force \textit{all} the $2N+1$ replica links (in both the bra and the ket) to have the same state thereby making them symmetric under the
permutations in
$S_{2N+1} \times S_{2N+1}$. Therefore, the full stat-mech model for this version of the MIPT is symmetric under $S_{2N+1}\times S_{2N+1}$ in the bulk.\\

While there is an enlarged permutation symmetry in the bulk, this symmetry is broken down at the top boundary, see  symmetry discussion at the end of section~\ref{globaltrmipt}. This is because the seam breaks $S_{2N+1} \times S_{2N+1}$ down to $H_N\times H_N\subset S_{2N+1}\times S_{2N+1}$. One can check this by considering the $N$-fold replicated ``seam" state  from \eqref{LabelEqSeamState} and re-writing it as 
\begin{multline}
(\langle \langle {\rm {\scriptstyle Seam}} |)^{\otimes N} = \sum_{\vec{\mathbf{i}}, \vec{\mathbf{k}}^*}
\langle\langle i_1, k^*_1, i_2,k_2,\dots,i_N,k_N^*| \otimes \\\langle\langle i_1, k_1^*, i_2,k_2^*,\dots,i_N,k_N^*|,\end{multline}
where $\vec{\mathbf{i}},\vec{\mathbf{k}} = (i_1,i_2,\dots,i_N),(k_1,k_2,\dots,k_N)$. 
The state above has a $Z_2\times Z_2$ symmetry built in, implemented by exchanging mirrored pair $i_n\leftrightarrow i_n, 1\leq n\leq N$ (and correspondingly for $k^*$). It is additionally 
invariant under the transpositions that exchange the pairs $(i_n, i_n) \leftrightarrow (i_m,i_m), 1\leq m,n \leq N$ (similarly for $k^*$), where the two $i_n's$ correspond to mirrored partners~\footnote{ in the notation of the second paragraph of Sec.~\ref{LabelSubSubSectionLocalTRModel}}. These are precisely the generators of $H_N\times H_N$ as explained at the beginning of section \ref{LabelSubSubSectionLocalTRModel}. This fact can also be seen graphically from figure \ref{stat_mech_model_mirror}. Specifically, if the replicas are labeled as 
$(1, 2, ..., N, \bar{1}, \bar{2}, ..., \bar{N})$, 
where the overbar denotes the mirrored counterpart of the replica 
as in Fig.~\ref{stat_mech_model_mirror}, the relevant $H_N$ subgroup
that
keeps the seam intact is generated by single transpositions $(l\; \bar{l})$ with $1 \leq l  \leq N $,
and double transpositions
$(l\; m)(\bar{l}\; \bar{m})$ with $1 \leq l, m 
\leq N$.
The single transpositions represent the additional non-permutation symmetry introduced by TR. Additionally, note that the Born-probability replica is excluded when discussing the $H_N$ subgroup, as it does not double and therefore does not have a mirrored partner.

While the local TR MIPT falls into the universality class of the prototypical Haar MIPT due to its full enlarged permutation symmetry $S_{2N+1}\times S_{2N+1}$, global TR gives us a new universality class of dynamics.
Note that this would have been the case even if we did not have local TR dynamics. This is because the permutation symmetry $S_{2N+1}$ emerges from the exchange of replicas - whose number increases from $N+1$ to $2N+1$ due to the mirrored structure enforced by global TR -  not from the presence of local TR symmetry. This establishes a key result of our work, namely, that ensemble averaged global TR symmetric monitored dynamics (with measurements post-selected to satisfy TR invariance in a ``strong'' way, {\it i.e.} in every quantum trajectory) represents a novel universality class of dynamics.\\

The experimental overhead of post-selecting in the global TR model is
in addition to the already exponential overhead of post-selecting to observe MIPTs, a problem famously 
termed
the `post-selection problem'~\cite{hoke_measurement-induced_2023}. Although there are several approaches to overcome this problem \cite{PhysRevLett.125.070606,noel_measurement-induced_2022, dehghani_neural-network_2023, PhysRevX.13.021026, lee2022decodingmeasurementpreparedquantumphases, PhysRevLett.129.200602, PRXQuantum.5.030311}, one would like to reduce the extent to which there is post-selection. This leads us to investigate the universality class of a model that mirrors the gates and the measurement locations but not the measurement outcomes. We shall subsequently refer to this model as the ``global TR symmetric model without post-selection", although it is important to keep in mind that global TR symmetry is microscopically broken when we do not post-select in the second half. A reason one can still however hope for the dynamics to fall in the same universality class as global TR dynamics is due to averaging occasionally giving rise to emergent symmetries. For example, Haar random circuits break translational symmetry yet the resultant stat-mech model is indeed invariant under translations. A similar emergence could potentially arise in the non-post-selected case, where one might hope that some notion of ``average TR-invariance'' emerges upon averaging over measurement outcomes.

It is easy to check whether such emergence occurs in this case. We consider the stat-mech model for the case with microscopic global TR symmetry and examine how the symmetries change when we do not post-select in the second half. Note that the measurements in the post-selected case respect $S_{2N+1}\times S_{2N+1}$ as explained earlier. In the non-post-selected case however,  only half the number $(N+1)$ of links in both the bra and ket are required to agree upon measurement since the latter half of the evolution now has a different trajectory. Consequently, any transformation that permutes between the two sets of $(N+1)$ links will generally not be a symmetry of the stat-mech model. Therefore, this model also has $S_{N+1}\times S_{N+1}$ symmetry, just like the local TR symmetric case. \\
To summarize, we have established that while local TR monitored dynamics falls into the (unitary)
Haar universality class, global TR dynamics (with post-selected measurements) presents a new universality class of dynamics. Although the symmetry properties are tractable, the stat-mech models \eqref{partitionbulkmiptnonmirrored} and \eqref{partitionfunctionmipt2global} are complicated due to their non-local link weights. The limit of large-$d$ however drastically simplifies the Boltzmann weights allowing us to make predictions about the nature of the entanglement transition \cite{PhysRevB.101.104302, PhysRevX.12.041002, PhysRevB.109.174307} as we shall now discuss.

\begin{figure*}[t!]
        \captionsetup{justification=Justified, singlelinecheck=false, font=small}
	\centering
	\includegraphics[width=\textwidth]{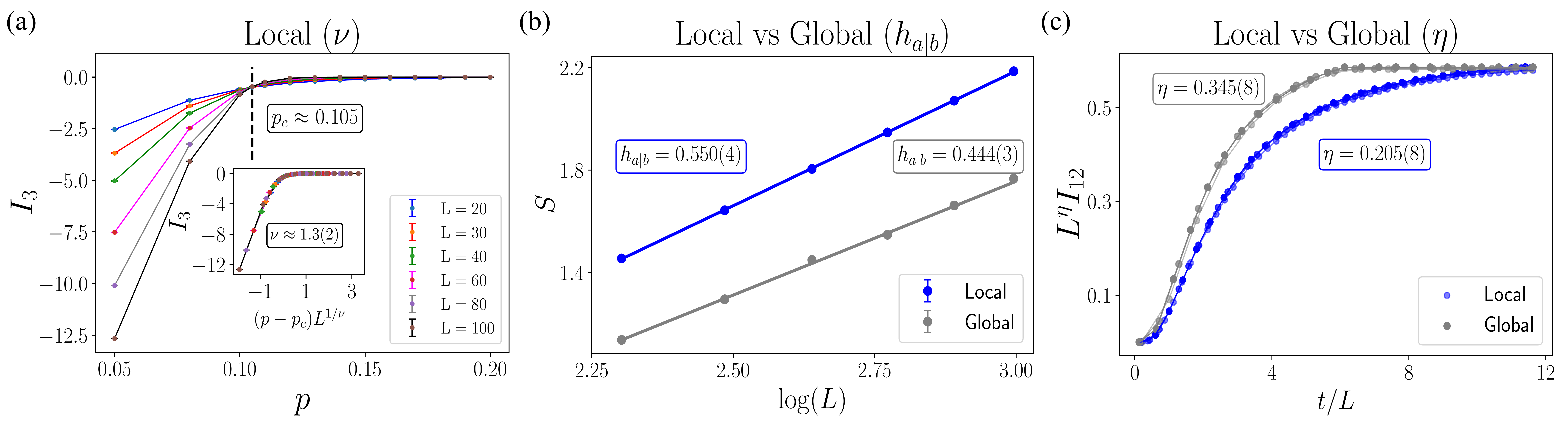}
	\caption{\textbf{TR-invariant Clifford numerics}: (a) $I_{3}$ vs $p$; We perform finite size scaling of $I_{3}$ to obtain the critical point $p_{c} \approx 0.105$ and correlation length exponent $\nu \approx 1.3(2)$ ( plotted in the inset) for the Local time-reversal model. We consider sizes $L= 20,30,40,60,80,100$, and average each data point over 50000 samples. (b) $S$ vs $L$; the bipartite entanglement entropy between two halves $A$ and $\bar{A}$ stays logarithmic ($\sim 2h_{a|b}\log(L)$) at the critical point $p_{c}$ ($p = 0.105$). We obtain the universal pre-factor of $\log$ at criticality from the linear fit between $S$ and $\log(L)$ . The blue and grey lines denote Local and Global, where the boundary exponent $h_{a|b} \approx 0.550(4)$ and $0.444(3)$ for Local and Global models, respectively. Clearly the values differ and confirm different universality classes of Local and Global. (c) $I_{12}(t,L)$ vs $t/L$; the dynamical scaling of space separated correlators $I_{12}(t,L)$ gives the bulk exponent $\eta$. We note a clear collapse and obtain bulk exponents $\eta =  0.205(8)$ and $0.345(8)$ for Local and Global models, respectively. Clearly, this again differs for the respective time-reversal models. We average each data over 100000 samples and show collapse for sizes $L = 10, 14,$ and $16$ in shades of blue and grey for Local and Global models, respectively.}
	\label{exponents_clifford}
 
\end{figure*}
\subsection{$d\;\rightarrow \infty\;$limit}
\label{large-d}

We begin by considering the simpler local TR symmetric model \eqref{partitionbulkmiptnonmirrored}. In the 
limit of infinite on-site Hilbert space dimension
$d \to \infty,$
the vertex weights of this stat-mech model simplify to \cite{matsumotogeneral}
\begin{equation}
\label{dinftylimitwg}
    \text{Wg}^{O(d^2+1)}(\tau) = d^{-2(N+1)}\delta_{\tau\in H_Q}
    + \mathcal{O}(d^{-2N-6}),
\end{equation}

where $\delta_{g\in H_{N+1}}$ is defined to be $=1$ only when $\tau\in H_{N+1}$, but $=0$ 
for any other elements $\tau\in S_{2(N+1)}$. This is based on the fact that the Weingarten function $\text{Wg}^{O(d^2+1)}(\tau)$
for $\tau\in S_{2(N+1)}$ is invariant under right- and left-multiplication of the argument $\tau$ by group elements of the 
subgroup $H_{N+1} \subset S_{2(N+1)}$ (as discussed, e.g., in  the second footnote in Sec.~\ref{LabelSubSubSectionLocalTRModel}; or see \cite{matsumotogeneral, collins_properties_2009}). Therefore, the Weingarten function is constant on the double cosets of $H_{N+1}$ in $S_{2(N+1)}$, and is only a function of the double cosets;
the subgroup $H_{N+1}$ is itself one of the double cosets, namely the ``identity double coset''~\footnote{the identity 
double coset is $H_{N+1}  e   H_{N+1}= H_{N+1}$ 
where $e=\text{identity}\in S_{2(N+1)}$, labeled by the 
partition $\lambda=(1,1,\dots,1)$}. [See (\ref{LabelEqDoubleCosetDecomposition}), Appendix \ref{H_Qgroup}.]
Unlike in the cases of group-based ensembles like Haar and Clifford for which the Weingarten functions attain a 
maximum
at a particular configuration (identity), the Weingarten functions here thus attain a 
maximum on
an entire double-coset, which is the ``identity double coset" = $H_{N+1}$.\\

As expected, we find that the $d\xrightarrow[]{}\infty$ limit of this model mimics that of the Haar MIPT stat-mech model \cite{PhysRevB.101.104302}. As we increase $d$, local loops begin to dominate thereby gradually erasing the distinction between local TR and the Haar MIPT models. This happens because local loops tend to increase the total number of loops in the configuration increasing their overall probability as per \eqref{partitionbulkmiptnonmirrored}. Locality implies we restrict to $S_{N+1} \subset H_{N+1}$ and the dominant configuration 
occurs
when we have $g_v = g_{v'}$, where $\{g_{v},g_{v'}\}\in S_{N+1}\subset H_{N+1}$, with $v,v'$ labeling all possible neighboring vertices. Note that this configuration also maximizes the vertex weights according to \eqref{dinftylimitwg}.\\
An advantage of a local Haar-like configuration is that we can now perform the average $\mathbb{E}_{\mathbf{X}}$ exactly as in the Haar case giving local link weights \cite{PhysRevB.101.104302}
\begin{equation}
\label{linkweightsdinfty}
    W_p(g_v,g_{v'}) = pd + (1-p)d^{N+1}\delta_{g_v,g_{v'}}.
\end{equation}
Equations \eqref{dinftylimitwg} and \eqref{linkweightsdinfty} collectively imply that the Boltzmann weights for the stat-mech model \eqref{partitionbulkmiptnonmirrored} reduce to 
those
of the Potts model with $(N+1)!$ colors \cite{PhysRevB.101.104302}.
In the replica limit $N\xrightarrow[]{}0$, we naturally obtain a percolation picture for this transition where have clusters of aligned spins diluted by measurements with the transition occurring at $p_c = 1/2$ above which the number of clusters proliferates.

The large-$d$ limit also allows us to take the replica limit \eqref{renyiavgreplica2} exactly and analytically realize the entanglement transition through the scaling of $\bar{S}_{n,A}$. The argument here is same as that in \cite{PhysRevX.12.041002}, where it was given in the context of a global $U(1)$ symmetry. We briefly repeat it here for completeness. Equation \eqref{renyiavgreplica2} can be re-written to express $\bar{S}_{n,A}$ as a difference in free energies $\Delta F = F_A - F_0 = -\log(Z_A/Z_0)$. In the limit $d\xrightarrow[]{}\infty$, $Z_A$ and $Z_0$ both have the same bulk configuration of aligned spins and only differ at the boundary with $Z_0$ having a uniform boundary and $Z_A$ having a domain wall due to a different boundary condition \eqref{boundarycondition2} in region A. The ratio $Z_A/Z_0 = d^{(k+N)l_{\rm{DW}}}$ (see \cite{PhysRevX.12.041002}), where $l_{\rm{DW}}=l_{\rm{DW}}(\mathbf{X})$ is the number of unmeasured links the DW crosses. Since the free energy cost $\Delta F\gg 0$, the DW will minimize $l_{\rm{DW}}$ following a ``minimal-cut" prescription. Finally, taking the replica limit gives ${S}_{n,A} = l_{\rm{DW}}(\mathbf{X})\log d$, valid for all values of $p$. Averaging over $\mathbf{X}$, one finds that for values of $p<1/2$, $l_{\rm{DW}}\sim L_A$, where $L_A$ is the length of the sub-system $A$. For $p>1/2$ on the other hand, $l_{\rm{DW}}\sim \mathcal{O}(1)$ with the transition occuring at $p_c = 1/2$ as predicted by percolation. \\

The above considerations do not change for the global TR scenario. The stat-mech model for the global TR MIPT has the same vertex and link weights (upto boundary conditions and difference in the number of replicas) due to which the large-$d$ limit here also mimics the Haar case and maps to percolation, this time in the $N\rightarrow 0$ limit of the Potts model $S_{(2N+1)!}$. One might initially expect the entanglement entropy that was realized analytically in the replica limit to also change by a factor because of the change the number of links at each site as compared to Haar/local TR. However, the mixed boundary conditions for global TR imply that only half of the first $2N$ links at the bottom boundary contribute to the partition function. Additionally, the extra replica continues to have identity ($e\in S_{2(N+1)}$) boundary conditions. These facts combined lead to the replica limit being unchanged from the Haar/local TR case.

\section{Numerics}

\label{numerics}
In this section, we present numerical evidence supporting the claims made in this work using Haar and Clifford monitored circuits. We demonstrate that the MIPT for global TR dynamics belongs to a new universality class. In contrast, the MIPT for local TR dynamics, as well as the non-post-selected version of global TR, share the same universality class. Additionally, for both the Haar and Clifford cases, we find that the universality classes of the latter align with those of the Haar and Clifford random MIPTs without global symmetries. \\

\subsection{Clifford TR Numerics}

The transition can be formulated using Clifford gates which respect time reversal symmetry. To construct such a gate set, we select unitaries satisfying $U = U^{T}$ from the finite two-qubit Clifford group. There are a total of 340 such gates, where elements show up from all four Clifford classes (namely, \rm{Single qubit}, \rm{SWAP}, \rm{iSWAP}, and \rm{CNOT} class) 
\cite{barends2014superconducting}. This indeed gives rise to non-trivial transition physics for Clifford time-reversal models. Moreover, the scalability allows to get precise estimates of universal exponents, which is further used to establish all claims made for universality classes of local and global time-reversal models.

\begin{table}[t!]
\captionsetup{justification=Justified, singlelinecheck=false, font=small}
\caption{\label{tab:table1}Critical exponents for the Local and Global time-reversal circuits at their respective critical point $p_{c}$. In particular, we extract the 
correlation length exponent $\nu$, boundary scaling dimension $h_{a|b}$, 
(typical) bulk exponent $\eta=2 x_1^{\rm typ}$,
and dynamical exponent $z$ for Clifford time-reversal circuits, and compute the effective central charge $c_{\rm{eff}}$ for the Haar time-reversal models.}

\begin{ruledtabular}
\begin{tabular}{ 
|p{1cm}||p{1.15cm}|p{1.15cm}|p{1.15cm}| }
\hline
 \multicolumn{1}{|c||}{} & \multicolumn{3}{c|}{Time-Reversal models}  \\
 \hline
 \multicolumn{1}{|c||}{Critical Parameters}& Local & $\rm{Global}^{*}$ & Global \\
 \hline
 \hline
 $p_{c} ~(\rm{TR-Clifford})$ & 0.105(1)& 0.105(2) & 0.120(3)  \\
 \hline
 $\nu$ & 1.3(2)& 1.1(1) & 1.0(3)  \\
 \hline
 $h_{a|b}$ & 0.550(4)& 0.550(4) & 0.444(3) \\   
 \hline
 $\eta$ & 0.205(8)& 0.215(6) & 0.345(8) \\  
 \hline
 $z$ & 1.06(8)& 1.02(6) & 1.00(9)   \\
 \hline
 \hline
 \hline
 $p_{c} ~(\rm{TR-Haar})$ & 0.15 & 0.15 & 0.146 \\ 
 \hline
 $c_{\rm{eff}}$ & 0.26(3)& 0.27(3) & 0.38(5) \\ 
 \hline
\end{tabular}
\footnotesize{$\rm{Global}^{*} \equiv $ Global TR model without post-selection in the second half.} \\
\end{ruledtabular}

\end{table}

We first obtain the critical point using standard MIPT order parameters such as tripartite mutual information $I_{3}$ (TMI) \cite{PhysRevX.10.041020,PhysRevB.101.060301} and ancilla order parameter $S_{q}$ \cite{PhysRevLett.125.070606}. We note a crossing and show a clear scaling of $I_{3}$ with $(p-p_{c})L^{1/\nu}$ in Fig.~\ref{exponents_clifford} (a) for the Local time-reversal model. This gives both $p_{c}$ and the correlation length exponent $\nu$. Similar scaling law allows to extract these values for other models which we list in Table.~\ref{tab:table1}. Note the values stay in agreement with the ancilla
probe (not shown here).
 
\begin{figure*}[t!]
	\centering
	\includegraphics[width=\textwidth]{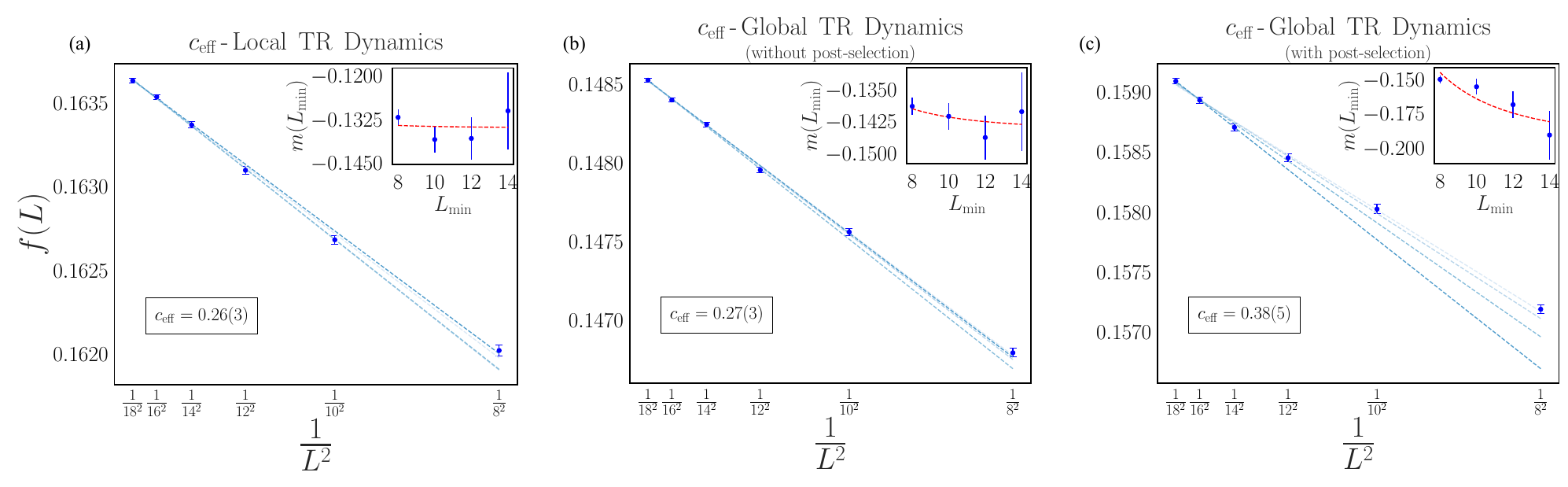}
	    \captionsetup{justification=Justified, singlelinecheck=false, font=small}
    \caption{\textbf{Free energy density $\mathbf{f(L)}$ vs $\mathbf{1/L^2}$ for Haar TR models:} The densities are plotted at respective $p_c$ as per figure \ref{tmi} for (a) Local TR dynamics (b) Global TR dynamics (c) Global TR dynamics without post-selection. The dotted lines indicate the linear fitting obtained using different $L_{\rm{min}}$. The inset shows resulting slopes $m_0(L_{\rm{min}})$ vs $L_{\rm{min}}$ where we omit the data-points for $L<L_{\rm{min}}$. We fit the slopes to the curve $m_0(L) = a + b/L^2$, allowing us to extract $c_{\rm{eff}}$ (see equation \eqref{freeenergydensity}). Each data point in the above plots is a result of averaging over more than $4 \times 10^4$ circuits.}
    \label{ceff_haar}
\end{figure*}
Now locating ourselves at $p_{c}$, we extract the standard boundary scaling dimension 
$h_{a|b}$~\cite{PhysRevB.104.104305,PhysRevB.109.014303,PhysRevB.109.174307} 
and
(typical) bulk exponent 
$\eta= 2 x_1^{\rm typ}$~\cite{PhysRevX.10.041020,PhysRevB.101.060301,PhysRevLett.125.070606,PhysRevLett.128.050602},
from $\log$ scaling of the
bipartite entanglement entropy $S$ and time dynamics of space separated 
(typical)
correlators $I_{12}(t,L)$, respectively. In particular, we show there comparison for the local and the global
model in Fig.~\ref{exponents_clifford}(b) and~\ref{exponents_clifford}(c). We go on computing similar exponents for the
non-postselected global time-reversal model, denoted as $\rm{Global}^{*}$. The values extracted for all cases are summarized in Table.~\ref{tab:table1}. The exponent values correctly identify that Local and Global belong to different universality  
classes,
whereas both Local and $\rm{Global}^{*}$ belong to the known Clifford universality class.  Additionally, we note for all models the dynamical exponent $z$ stays close to $1.0$.

\subsection{Haar TR Numerics}

For the Haar case, we start by locating $p_c$ using  TMI 
as a probe just like in the Clifford case. We find that the TMI records are remarkably similar across the three variants (see figure \ref{tmi} in appendix). Given such close entropy records, we anticipate that the correlation length exponents $\nu$ and the prefactor of the logarithm 
for the R\'enyi entropy $S_{n}$
at 
criticality~\footnote{
sometimes~\cite{PhysRevB.101.060301}
denoted by $\alpha(n)$, not to be confused with the anisotropy factor $\alpha$
(App.~\ref{anisotropycalc}),
and for the case $n=1$ 
typically~\cite{PhysRevB.104.104305,PhysRevB.109.174307, PhysRevB.109.014303}
by ${1\over 2}\cdot \alpha(1)=$ $h_{a|b}$.}
to be very similar across these models. Therefore, to effectively distinguish the universality classes of the Haar circuits, we focus on evaluating the effective central charge $c_{\rm{eff}}$ instead. This approach allows us to better characterize the critical properties between the Haar TR symmetric models that are not apparent from the TMI data alone.

\subsubsection{Effective Central Charge $(c_{\rm{eff}})$}
Before we discuss our results for Haar TR models, it is first worth reviewing how the effective central charge $c_{\rm{eff}}$ is described. Our discussion closely follows
\cite{PhysRevLett.128.050602}
and \cite{PhysRevB.109.014303}. \\

An alternate way to formulate the stat-mech models for these MIPTs is to think of each trajectory as a (1+1)-dimensional statistical mechanics model with the partition function $Z_{\mathbf{m}} \equiv p_{\mathbf{m}}$. The trajectories altogether form an ensemble of statistical mechanics models with quenched disorder due to the measurements, and where we weigh each trajectory with $p_{\mathbf{m}}$. The quenched free energy $F$ of the resultant stat-mech model can then be written as 
\begin{equation}
    F = -\sum_{\mathbf{m}}p_{\mathbf{m}}\log p_{\mathbf{m}}.
\end{equation}
$F$ is the quantity that we numerically calculate. Notably, the finite-size scaling of $F$ is governed by conformal invariance. To understand this, we can consider the partition function of the stat-mech model as a replicated partition function ${Z}^k_{\mathbf{m}} = p^k_{\mathbf{m}}$, weighted with the Born probability $p_{\mathbf{m}}$. This gives us $\bar{Z}_k =\sum_{\mathbf{m}}p^k_{\mathbf{m}} p_{\mathbf{m}}$, which is precisely the form of the partition functions of the stat-mech models we have discussed earlier \footnote{There is a minor but subtle change to this statement for the global post-selected scenario. Using equation \eqref{partitionfunctionmipt2global}, we see that the replicated annealed averaged partition function $\bar{Z}^k = \sum_{\mathbf{m}}p^k_{\mathbf{m}} q_{\mathbf{m}}$, where we use the born probability $q_{\mathbf{m}}\neq p_{\mathbf{m}}$ of only the first half of the trajectory to weigh each term in the partition function $Z^k_{\mathbf{m}}$. It follows from this that $F = -\sum_{\mathbf{m}}q_{\mathbf{m}}\log p_{\mathbf{m}}$ for the global post-selected case, which is \textit{not} the
(Shannon) entropy of the measurement record
as is the case for the local TR symmetric and Haar models. }. The corresponding annealed averaged free energy can be defined as $F_k = -\log \bar{Z}_k$. Using the replica trick, one can recover $F = \lim_{k\to 0}\frac{dF_k}{dk}$. Since we know that $F_k$ undergoes a second order phase transition governed by conformal invariance (with the replica limit coinciding with the MIPT of interest), $F$ does too. More precisely, standard results in CFT imply that the bulk free energy density $f(L) = F/(\alpha t L)$ at criticality scales as
\begin{equation}
\label{freeenergydensity}
    f(L) = f(L=\infty) - \frac{\pi c_{\rm{eff}}}{6L^2} + \dots,
\end{equation}
where we have considered a cylindrical geometry of circumference $L$, circuit depth $t$, and where $\alpha$ characterizes the asymmetry between space and time in our stat-mech model. Using the replica limit, the effective central charge above can be written as $c_{\rm{eff}} = \lim_{k\to 0}\frac{dc(k)}{dk}$, where $c(k)$ is the central charge for the model defined by $\bar{Z}^k$. It is important to note that $c(k)$ itself is not the quantity of interest. This is because $\lim_{k\to 0} c(k)=0$, due to the partition function $\bar{Z}^k$ becoming trivial in this limit (i.e., $\lim_{k\to 0} \bar{Z}^k=1$). \\

\subsubsection{Results}
Here we present the results for the Haar TR circuits. The free energy calculations were performed following the procedure outlined in the supplemental material of \cite{PhysRevLett.128.050602}. The protocol for evaluating the anisotropy parameter for the local TR case is also the one used in \cite{PhysRevLett.128.050602}. However, for global TR-symmetric models, the protocol was modified to account for the two-to-one space-time correspondence between the circuit and the stat-mech model (see figure \ref{stat_mech_model_mirror}). Full details of this adjustment can be found in Appendix \ref{anisotropycalc}.

The anisotropy values obtained are $\alpha_{\rm{global}} = 0.61(8)$, $\alpha_{\rm{global*}} = 0.64(6)$, and $\alpha_{\rm{local}} = 0.59(5)$. Figure \ref{ceff_haar} shows the free energy density results, where $f(L)$ is plotted against $1/L^2$. As expected, the behavior is linear in all cases. Using a double fitting procedure \cite{PhysRevLett.128.050602, PhysRevB.109.014303}, we estimate $f^{\mathrm{global}}(L) = -0.19(3) + \frac{3.42(1)}{L^2}$, $f^{\mathrm{local}}(L) = -0.13(1) + \frac{0.20(2)}{L^2}$, and $f^{\mathrm{global*}}(L) = -0.14(3) + \frac{0.25(3)}{L^2}$.

We then compute the effective central charges using \eqref{freeenergydensity}. For the local TR dynamics and global TR dynamics without post-selection, the effective central charges indeed match that of  Haar ($c_{\rm{eff}} = 0.25(3)$), yielding $0.26(3)$ and $0.27(3)$, respectively. Furthermore, we find $c_{\rm{eff}} = 0.38(5)$ for the global TR dynamics (with post-selection), placing it in a new and different universality class as claimed earlier in this work. 
\section{Outlook}
\label{outlook}
Our work explores time-reversal (TR) symmetry in chaotic quantum systems through the framework of random quantum circuits. We distinguish between two types of TR symmetry in these circuits: local and global. Local TR symmetry requires each individual gate $U$ to be symmetric ($U^T = U$), while global TR symmetry requires that the full evolution is of that form. We show that this amounts to picking gates from the Circular Orthogonal Ensemble (COE).
By averaging over moments of the COE, we then develop a replica stat-mech model generalizing the approach of refs. ~\cite{PhysRevB.99.174205, PhysRevB.101.104301, PhysRevB.101.104302}. This mapping could then be
employed to probe ensemble-averaged values of observables such as Rényi entropies and characterize operator spreading through calculation of out-of-time-ordered correlators and other many-body quantum chaos metrics. As an example of  application, we used this model to investigate MIPTs in monitored TR-invariant quantum systems, where the measurements are performed in a TR invariant basis.\\

Despite making measurements in the TR invariant basis, our symmetry analysis reveals that the local TR symmetric model shares the same $S_N \times S_N$ symmetry as the Haar stat-mech model (where $N$ is the number of replicas), 
suggesting that they belong to the same universality class. In contrast, the global TR case, owing to the folded geometry of its stat-mech model (see figure \ref{stat_mech_model_mirror}), displays an enlarged replica permutation symmetry, placing it in a new universality class. We find that this new universality class is sensitive to the microscopic global TR symmetry. Specifically, not post-selecting measurements while maintaining a globally symmetric structure for the unitary part (while also making measurements symmetrically in a TR basis) reverts the universality class back to that of Haar. In this sense, there is no emergent TR symmetry in the global TR stat-mech model when we do not post-select the outcomes in the second half of the time-evolution.

We present numerical evidence to substantiate the above claims by analyzing using both Haar and Clifford monitored circuits. Specifically, we observe that the Clifford variant and the Haar variant of the global TR symmetric model exhibit distinct critical exponents and 
effective central charges respectively. In contrast, both the non-post-selected global model and the local TR symmetric model, which share the same symmetries, are shown to have nearly identical critical exponents and central effective charges that differ from the global post-selected case.\\

In this work we have addressed only the case of time-reversal invariance where the time-reversal operator squares to plus the identity,
${\cal T}^2=+1$ (``orthogonal symmetry class''). Details of extension, e.g., to the case where ${\cal T}^2 = -1$ (``symplectic symmetry class'') will be reserved for future work.\\

\textbf { Acknowledgments -- } This work was supported by the US Department of Energy, Office of Science, Basic Energy Sciences, under award No. DE-SC0023999.  R.V. acknowledge hospitality of KITP during the DYNISQ22 follow-on program ``Phases of active quantum matter'' during which parts of this work were completed. KITP is supported by grant NSF PHY-2309135. We thank Chao-Ming 
Jian
and Andrew Potter for useful discussions.

\appendix
\section {(Double) Coset Types of $H_N\subset S_{2N}$}
\label{H_Qgroup}
Consider the subgroup $H_N\subset S_{2N}$ 
generated by the transpositions 
$(l\; l'),1\leq l \leq N$, and 
double transpositions $(l\; m), (l'\;m'), 
1\leq l<m\leq N
$. Let $\sigma\in S_{2N}$ and create an un-directed graph with vertices $\{1,\dots, 2N\}$ and edges between the vertex pairs $\{l\; l'\}$, $\{\sigma(m),\sigma(m')\}$. Each vertex is then part of two edges and the number of vertices in each connected component is even. Let them be indexed as $2\lambda_1\geq 2\lambda_2\geq\dots2\lambda_l$ (where we have $l$ connected components). Then $\lambda=(\lambda_1,\lambda_2,\dots\lambda_l)$ is a partition of $N$
(in symbols $\lambda \vdash N$), and one defines $\lambda$ to be the {\it ``coset type'' of} $\sigma$. Note that this is exactly how the contraction in equation \eqref{eq:inlinecoecalculus3} was defined as well. Additionally, one can check that given $\sigma,\tau \in S_{2N}$, their coset types coincide iff 
$H_N\sigma H_N = H_N \tau H_N$
resulting in a double coset decomposition of
$S_{2N}$ with respect to the subgroup
$H_N \subset S_{2N}$,
\begin{equation}
\label{LabelEqDoubleCosetDecomposition}
    S_{2N} = \bigsqcup_{k\vdash N} H_k,
\end{equation}
where $H_k :=\{\sigma \in S_{2N}\;|\;\text{the coset type of }\sigma \text{ is } k\}$. Finally, using the definition of $H_N$ in \ref{universality}, it is easy to show
that $H_N:=H_{\{1\}^N}$, which is the double coset of the identity permutation $e\in S_{2N}$: $H_{N}=H_N e H_N$.\\

\begin{figure*}[t!]
	\centering
	\includegraphics[width=\textwidth]{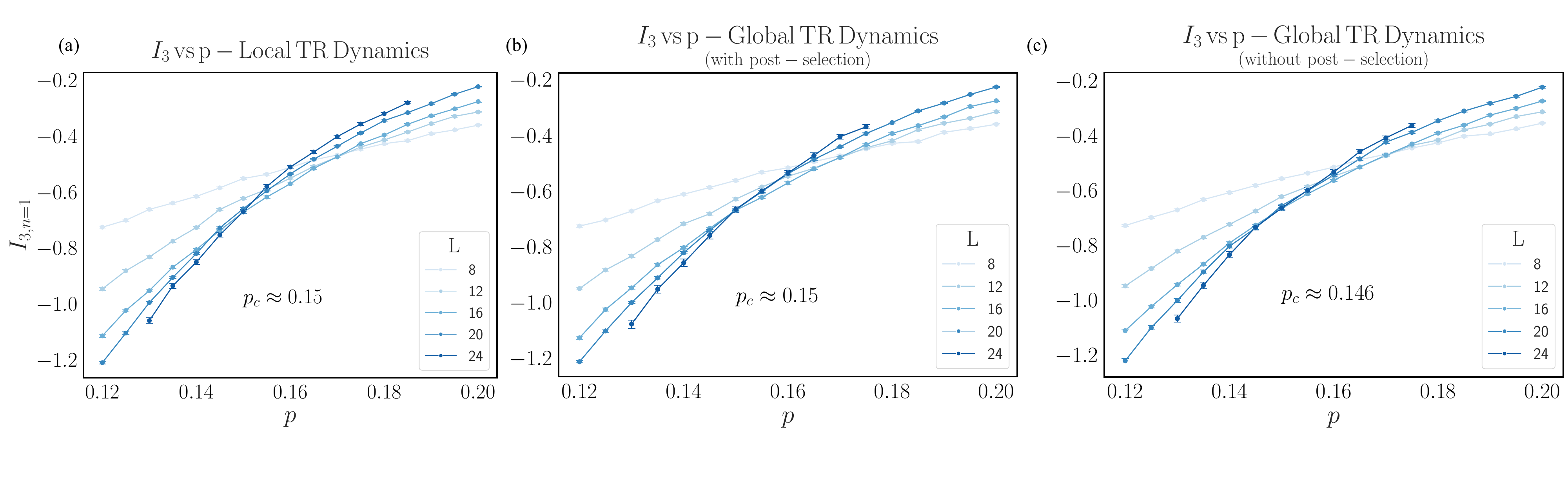}
	    \captionsetup{justification=Justified, singlelinecheck=false, font=small}
    \caption{\textbf{Tripartite Mutual Information (TMI) vs p for $\mathbf{L = 8,12,16,20,24}$ for Haar TR models}. In all three cases, each point above is a result of averaging over $10,000$ circuits for $L=8,12,16,20$ and $2,000$ circuits for $L=24$. (a) Global TR with $p_c \approx 0.15$. (b) Local TR with $p_c \approx 0.15$. (c) Global TR without post-selection with $p_c \approx 0.146$ .}
    \label{tmi}
\end{figure*}

\section{A Protocol to Measure the Anisotropy Parameter}
\label{anisotropycalc}

We present the numerical protocols used to evaluate the anisotropy factor for various stat-mech models introduced in the main text. We begin with a brief review of the protocol outlined in the Appendix of ref. \cite{PhysRevLett.128.050602}, which we use for the local TR symmetric case. We then discuss a modified protocol that we introduce for the global TR symmetric stat-mech models (both post-selected and non-post-selected versions).\\
The free energy density $f = F/A$ is measured per unit spacetime area $A = \alpha Lt$, where $\alpha$ measures the asymmetry between space and time with $L = \alpha t$. To measure this asymmetry, we compare correlation functions in the space and time directions at criticality. Due to conformal invariance, we can map correlation functions on our (1+1)-d stat-mech models onto correlation functions on a cylinder using the conformal mapping $z' = f(z) = \frac{L}{2\pi}\ln z$. This results in the correlation function between two points $g'(z'_1,z'_2)$ on the cylinder:
\begin{equation}
    g'(z'_1,z'_2) = \bigg(\frac{\pi}{L}\bigg)^{2\Delta} \frac{1}{\abs{\sin [\frac{\pi}{L}(z_1'-z_2')]}^{2\Delta}},
\end{equation}
where $\Delta$ is the conformal dimension. We can extract the anisotropy factor by setting $g'_{\rm{time}}:= g(0,\alpha t)$ and $g'_{\rm{space}}:=g(0,iL/2)$. This gives
\begin{equation}
\frac{g'_{\rm{time}}}{g'_{\rm{space}}} = \bigg(\frac{2e^{\pi \alpha t/L}}{e^{2\pi\alpha t/L}-1}\bigg)^{2\Delta}.
\end{equation}
Setting $\frac{g'_{\rm{time}}}{g'_{\rm{space}}}=1$ for some $t=t_*$ eliminates the dependence on $\Delta$ to give
\begin{equation}
\label{alphaeq}
    \alpha = \ln(1+\sqrt{2})\frac{L}{\pi t_*}.
\end{equation}
To implement the above numerically, we first run the dynamics until $\tau_1 = 4L$ to allow the system to equilibrate. We then measure the qubit at site $x_1$ and Bell entangle an ancilla qubit to it. Following that, we run the dynamics up until $\tau_2 = \tau_1 + \delta \tau$, measure at site $x_2 = x_1 + \delta x$, and similarly entangle another ancilla to this site. We then follow the mutual information between the ancillas to obtain $I_{12}(\delta x, \delta \tau)$. We use $\delta x = 0$ to obtain $g'_{\mathrm{time}}$ for various $\delta \tau$, and $\delta \tau = 0$, $\delta x = L/2$ to obtain $g'_{\mathrm{space}}$.\\
The aforementioned protocol largely holds true but needs careful modification to apply to the case of the mirrored stat-mech model introduced in section \ref{globaltrmipt}. There are two considerations that modify our protocol. Firstly, note that time in the stat-mech model corresponds to a vertical interval in figure \ref{stat_mech_model_mirror}, which is not the same as a time measured by the time steps of a circuit. Secondly, since a given spacetime point on the stat-mech model corresponds to two instances in the real-time circuit, extracting any correlation function in the stat-mech picture must involve performing an operation (such as measuring and entangling with an ancilla) at \textit{both} a given site and its mirrored counterpart. In other words, our protocol must necessarily respect the mirror symmetry of the model while implementing the standard protocol. Here is the step-by-step procedure we use in order to achieve the above:
\begin{figure*}[t!]
	\centering
	\includegraphics[width=\textwidth]{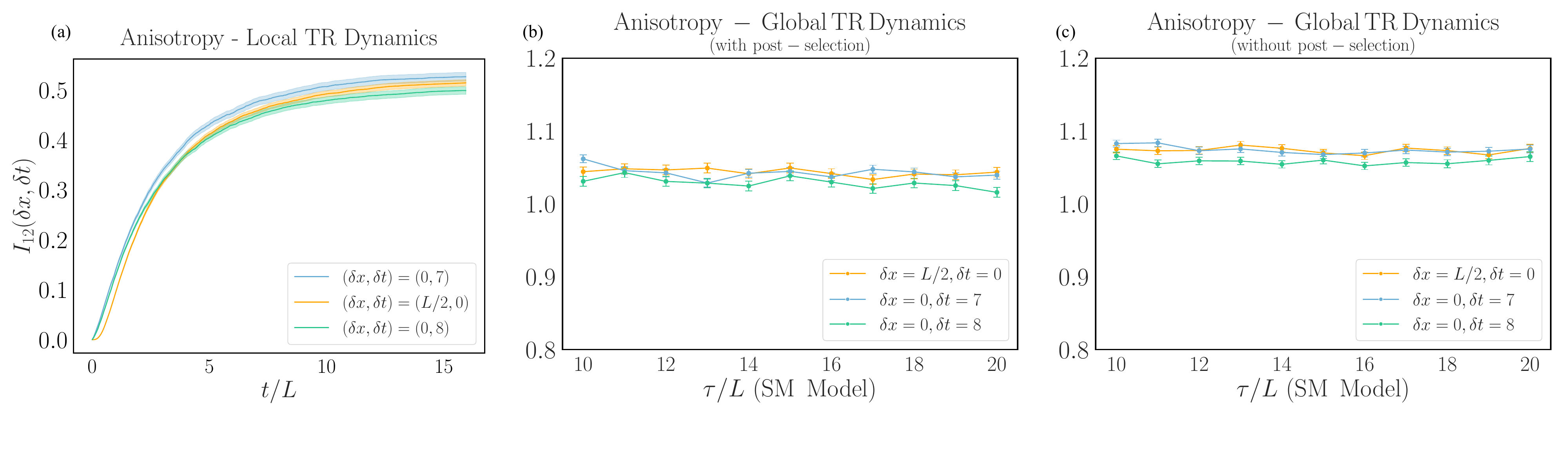}
	    \captionsetup{justification=Justified, singlelinecheck=false, font=small}
    \caption{\textbf{Space and time correlation functions for calculating anisotropy for Haar TR models:} These are plotted at $p_c$ of various TR symmetric models discussed in the main text for size $L=16$ (a) Local TR dynamics. Each point above is sampled from over 3000 circuits. (b) Global TR dynamics without post-selection. Each point is sampled from over 20000 circuits. (c) Global TR dynamics with post-selection. Each point above is sampled from over 20000 circuits.}
    \label{anisotropy_plots}
\end{figure*}
\begin{enumerate}
    \item We run the dynamics up until $\tau_1 = 20L$, after which we measure (with outcome $m_1 \in \{0, 1\}$) and entangle ancilla 1 at site $x_1$.
    \item We further run the dynamics until time $\tau_2 = \tau_1 + \delta \tau$, measure the qubit at site $x_2$ (with outcome $m_2 \in \{0, 1\}$), and entangle a new ancilla 2 to this site.
    \item We follow the dynamics for $\tau$ timesteps, completing the first half of our evolution.
    \item We begin the mirrored evolution for $\tau$ timesteps, using identical gates, measurement locations, and post-selecting outcomes to the corresponding first half of the evolution. We skip the post-selection of outcomes for the non-post-selected global TR model as discussed in section \ref{globaltrmipt}. 
    \item We store the qubit at site $x_2$ and replace it with a new qubit at this site in state $\ket{m_2}$. This is the mirror operation of measuring at site $x_2$ and bell-entangling it with an ancilla. One can think of the stored qubit as a mirrored ancilla and the new replaced qubit as mirror of the measured qubit in the forward evolution.
    \item We evolve the system for time $\delta \tau$ and perform the same procedure of storing the qubit at site $x_1$ and replacing it with a new qubit in the state $\ket{m_1}$.
    \item We follow the mirrored dynamics for the rest of the time ($20L$) while keeping track of the mutual information $I_{12}(\delta x, \delta \tau)$ between the systems $1 \equiv \{\text{qubit at site $x_1$ + ancilla 1}\}$ and $2 \equiv \{\text{qubit at site $x_2$ + ancilla 2}\}$.
\end{enumerate}
It is important to note that $\tau$ introduced above is the time in our stat-mech model (as shown in figure \ref{anisotropy_plots}b and \ref{anisotropy_plots}c), and it is with respect to this time that we must record $I_{12}$. In particular, this implies that in order to get $I_{12}$ vs $\tau$, one must repeat the above procedure for different values of $\tau$. \\

\section{$p_c$ and Anisotropy Factors for Haar Random TR symmetric Models}
\subsection{$p_c$
for Haar TR Models}
We obtain the critical point for the Haar random TR models using TMI $I_{3,n=1}$ as a probe \cite{PhysRevB.101.060301, PhysRevX.10.041020}. TMI is defined as 
\begin{multline}
I_{3,n} = S_{n}(A)+S_{n}(B) + S_{n}(C) - S_{n}(A\cup B)\\ - S_{n}(B\cup C) - S_{n}(C\cup A) + S_{n}(A\cup B\cup C).
\end{multline}
where $S_n(A)$ is the subsystem Rényi entropy defined in \eqref{renyientropy}, and regions A, B, and C are consecutive sections of the chain of length $L/4$, where $L$ is the length of the full chain. The resulting plots are shown in figure \ref{tmi}.\\

\subsection{Anisotropy for Haar TR Models}
The result of the above procedures is displayed in figure \ref{anisotropy_plots}. We obtain records of $g_{\rm{space}}(\delta x =N/2)$ and $g_{\rm{time}}(\delta \tau)$ at their respective $p_c$ so that $g_{\rm{time}}(\delta \tau )\leq g_{\rm{space}}\leq g_{\rm{time}}(\delta \tau +1 )$. We then linearly interpolate to find the matching time $t_{\star}$ and use \eqref{alphaeq} to obtain the anisotropy factor \cite{PhysRevLett.128.050602}.  For local TR dynamics, we estimate the matching time $t_{*}\approx 7.51$, which results in $\alpha_{\rm{local}} \approx 0.59(5)$. For the global TR case (with post-selection), we observe that the correlation functions settle at $\tau \approx 10L$, where $\tau$ is time in the statistical mechanics model. We estimate $\alpha_{\rm{global}}$ as the average of the anisotropy factors obtained from each of the data points in figure  \ref{anisotropy_plots}b, resulting in $\alpha_{\rm{global}} \approx 0.61(8)$. In the same manner we obtain, resulting in $\alpha_{\rm{global*}} \approx 0.64(6)$ for the non-post-selected version of the global TR model. 

\nocite{*}

\bibliography{apssamp}

\end{document}